\documentclass{aa}  
\usepackage{graphicx}
\usepackage[colorlinks=true,citecolor=blue]{hyperref}
\usepackage{multirow}

\usepackage{wasysym}
\newcommand{\ASOHF}[0]{\texttt{ASOHF} }
\newcommand{\ASOHFns}[0]{\texttt{ASOHF}}
\newcommand{\dv}[2]{\frac{\mathrm{d}#1}{\mathrm{d}#2}}
\newcommand{\dd}[0]{\mathrm{d}}

\usepackage{xcolor}

\usepackage{txfonts}

\newcommand{\unit}[1]{\,\mathrm{#1}}

\begin{document}

   \title{The halo finding problem revisited: a deep revision of the ASOHF code}


   \author{David Vallés-Pérez
          \inst{1}\fnmsep\thanks{\email{david.valles-perez@uv.es}}
          \and
          Susana Planelles\inst{1,2}
          \and 
          Vicent Quilis\inst{1,2}
          }

   \institute{Departament d’Astronomia i Astrofísica, Universitat de València, E-46100 Burjassot (València), Spain
              \and
              Observatori Astronòmic, Universitat de València, E-46980 Paterna (València), Spain}

   \date{\today}

  \abstract
   {New-generation cosmological simulations are providing huge amounts of data, whose analysis becomes itself  a cutting-edge computational problem. In particular, the identification of gravitationally bound structures, known as halo finding, is one of the main analyses. A handful of codes developed to tackle this task have been presented during the last years.}
   {We present a deep revision of the already existing code \ASOHFns. The algorithm has been throughfully redesigned in order to improve its capabilities to find bound structures and substructures, both using dark matter particles and stars, its parallel performance, and its abilities to handle simulation outputs with vast amounts of particles. This upgraded version of \ASOHF is conceived to be a publicly available tool.}
   {A battery of idealised and realistic tests are presented in order to assess the performance of the new version of the halo finder. }
   {In the idealised tests, \ASOHF produces excellent results, being able to find virtually all the structures and substructures placed within the computational domain. When applied to realistic data from simulations, the performance of our finder is fully consistent with the results from other commonly used halo finders, with remarkable performance in substructure detection. Besides, \ASOHF turns out to be extremely efficient in terms of computational cost.}
   {We present a public, deeply revised version of the \ASOHF halo finder. The new version of the code produces remarkable results finding haloes and subhaloes in cosmological simulations, with an excellent parallel performance and with a contained computational cost.}


   \maketitle

\section{Introduction}
\label{s:intro}

Over the last four decades, numerical simulations of cosmic structure formation have grown significantly, both in size, dynamical range and accuracy of the physical model ingredients \citep[see, e.g.,][for recent reviews]{Vogelsberger_2020, Angulo_2022}. Besides a precise description of the evolution of the dark matter (DM) component of the Universe, current cosmological simulations have also improved their modeling of the complex baryonic physical processes shaping the properties of the gaseous and the stellar components \citep[see, e.g.,][for a review]{Planelles_2015}. On the other hand, the major development of computing facilities has led to an increasing computational power and to important advances in algorithms and techniques. This progress has allowed current cosmological simulations to reach a significant level of mass and force resolution, complexity and realism. 

Nowadays, the excellent predictive power of these simulations makes them essential tools in Cosmology and Astrophysics: they are crucial, not only to test the accepted cosmological paradigm, but to interpret and analyse how different physical processes inherent to the cosmic evolution affect the observational properties of the Universe we inhabit. To properly exploit the unprecedented capabilities of current state-of-the-art cosmological simulations, equivalently complex and sophisticated structure finding algorithms are also required. Indeed, a proper identification and characterization of the population of DM haloes and subhaloes, including their abundances, shapes and structure, physical properties and merging histories, turns out to be decisive to understand the formation and evolution of cosmic structures.

Even when a DM halo is `simply' a locally overdense, gravitationally bound structure embedded within the global background density field, there is an  unavoidable arbitrariness in the definition of its boundary and, hence, on its mass. The situation is even more accentuated in the case of subhaloes located within larger-scale overdensities, called hosts. In the last decades, in an attempt to overcome these issues, a significant number of halo-finding methods and techniques have arisen (e.g., \texttt{FoF}, \citealp{Davis_1985}; \texttt{SO}, \citealp{Cole_1996}; \texttt{HOP}, \citealp{Eisenstein_1998}; \texttt{BDM}, \citealp{Klypin_1999}; \texttt{Subfind}, \citealp{Springel_2001}, \citealp{Dolag_2009}; \texttt{AHF}, \citealp{Gill_2004, Knollmann_2009}; \texttt{ASOHF}, \citealp{Planelles_2010}; \texttt{Velociraptor}, \citealp{Elahi_2011}, \citealp{Elahi_2019}; \texttt{HBT}, \citealp{Han_2012}, \citealp{Han_2018}; \texttt{Rockstar}, \citealp{Rockstar_2013}; to cite a few).

In general, however, all these algorithms can be broadly divided on three main families of codes:  those based on the spherical overdensity method \citep[SO;][]{Press_1974, Cole_1996}, those relying on the friends-of-friends algorithm in 3D configuration space \citep[FoF;][]{Davis_1985}, and those others based on the 6D phase-space FoF \citep[e.g.][]{Diemand_2006}. Algorithms in the first class involve locating peaks in the density field, and finding the extent of the halo, either by growing spheres until either the enclosed density falls below a given threshold, or other properties are met. On the other hand, codes in the second and third categories link particles which are close to each other, either in configuration or in phase-space. In \cite{Knebe_2011, Knebe_2013} an exhaustive review and comparison of some of these halo finders was performed. These studies shown that, while all codes are able to identify the location of mock, isolated haloes, and some of their properties (e.g., the maximum circular velocity, $v_\mathrm{max}$), important differences arose when dealing with small-scale haloes, especially substructures \citep{Onions_2012, Onions_2013, Hoffmann_2014}, or when determining some properties such as spin and shape.

Besides DM structures, modern simulations also track the formation of stellar particles from cold gas, which cluster and form stellar haloes, or galaxies. While the formation of galaxies is tightly linked to their underlying DM haloes, their different evolutionary histories, properties and dynamics make them deserve to be studied as objects in their own right. Consistently, specific algorithms for such task have been developed recently \citep[e.g.,][]{Navarro_2013, Canas_2019}.

In \cite{Planelles_2010} we presented \ASOHFns, a halo finder based on the SO approach. Although \ASOHF\ was especially designed to be applied on the outcomes of grid-based cosmological simulations, it was adapted to work as a stand-alone halo finder on particle-based simulations. The performance of \ASOHF\ in different scenarios was demonstrated in \cite{Knebe_2011}. In addition, the code has been employed in a number of works \citep{Planelles_2013, Quilis_2017, Martin-Alvarez_2017, Planelles_2018, Valles_2020, Valles_2021}. The incessant improvements of cosmological simulations, aided by the ever-growing computing power available, have enormously increased the amount of particles and, consequently, the richness of small-scale structures. This demanded to revisit the process of halo-finding in \ASOHFns, both to make sure that small structures are well captured and their properties can be recovered unbiased, and to guarantee that the code is able to tackle these amounts of data within reasonable computational resources. 

In this paper we present  an upgraded, faster and more memory-efficient version of \ASOHF that is capable of efficiently dealing with the new generation of cosmological simulations which include huge amounts of DM particles, haloes and substructures. Amongst the main improvements, we find a smoother density interpolation which lowers the computational cost by decreasing the number of spurious density peaks, the addition of complementary unbinding procedures, a new scheme for looking for substructures, the ability to identify and characterise stellar haloes, and a domain decomposition approach which can lower the computational cost and computing time. On the performance side, the code has experienced a profound overhaul in terms of parallelisation and memory requirements, being able to analise simulations with hundreds of millions of particles on desktop workstations within a few minutes, at most. This version of \ASOHF is publicly available.

The paper is organized as follows. In Sect. \ref{s:algorithm} we describe the main procedure on which our halo finder relies to identify the samples of haloes, subhaloes and stellar haloes, as well as additional features such as the domain decomposition scheme, the merger tree, etc. The performance and scalability of \ASOHF\ in some idealised, yet rather complex, tests is shown in Sect. \ref{s:mock_tests}. Additionally, in Sect. \ref{s:simulation_tests} we test the performance of the code against actual simulation data and compare its performance to other well-known halo finders, and we show the capabilities of \ASOHF as a stellar halo finder. Finally, in Sect. \ref{s:conclusions} we discuss and summarise our results. Appendix \ref{s.appendix.unbinding} further describes one of the unbinding schemes, while in Appendix \ref{s.appendix.gravitational_energy} we discuss our estimation of the gravitational binding energy and most-bound particle by sampling.

\section{Algorithm}
\label{s:algorithm}
While the original algorithm was introduced by \cite{Planelles_2010}, a large number of modifications and upgrades have been undertaken in order to provide a fast, memory-efficient and flexible code which is able to tackle a new generation of cosmological simulations, with several orders of magnitude increase in the number of dark matter particles, haloes and a rich amount of substructure. Here we describe the main steps of our halo finding procedure. The implementation of \ASOHF for shared-memory platforms (OpenMP), written in \texttt{Fortran}, is publicly available through the GitHub repository of the code\footnote{\url{https://github.com/dvallesp/ASOHF}}. The following subsections describe the input data (Sect. \ref{s:algorithm.io}), the process of identification of density peaks (Sect. \ref{s:algorithm.grid}), the characterisation of haloes using particles (Sect. \ref{s:algorithm.particles}), the substructure identification scheme (Sect. \ref{s:algorithm.substructure}), the characterisation of stellar haloes (Sect. \ref{s:algorithm.stellar}), and several additional features and tools of the \ASOHF package (Sect. \ref{s:algorithm.additional}). Table \ref{tab:parameters} contains a summary of the configurable parameters of the code.

\begin{table*}[]
    \centering
    \caption{Summary of the main parameters that can be tuned to run \ASOHFns. The first block contains the parameters that have effect on the identification of haloes, while the second block refers to the stellar halo finding procedure.}
    \begin{tabular}{ll|l}
         Parameter & (Symbol) & Description and remarks  \\ 
         \hline
         \multicolumn{3}{c}{Parameters for halo finding: mesh creation and halo identification} \\ \hline
         Base grid size & $N_x$ & Typically set to $N_x = \sqrt[3]{N_\mathrm{part}}$ or $2\sqrt[3]{N_\mathrm{part}}$ \\
         Number of refinement levels & $n_\ell$ & Peak resolution will be $L/(N_x\times 2^{n_\ell})$, typically set to match \\
         & & the force resolution of the simulation \\
         Number of particles to flag a cell as `refinable' & $n_\mathrm{part}^\mathrm{refine}$ & \\
         Fraction of refinable cells to extend the patch & $f_\mathrm{refinable}^\mathrm{extend}$ & A patch is grown along a direction only if more than this  \\
         & & fraction of the newly added cells is refinable \\
         Minimum size of the patch to be accepted & $N_\mathrm{min}^\mathrm{patch}$ & Only patches with minimum dimension above this are accepted\\
         \multicolumn{2}{l|}{Base grid refinement border} & Exclude these many cells close to the domain boundary ($\geq 1$)\\
         \multicolumn{2}{l|}{AMR grids refinement border} & Idem., in each AMR grid ($\geq 0$) \\
         \multicolumn{2}{l|}{Kernel order for interpolating density from particles} & Either 1 (linear kernel) or 2 (quadratic kernel; recommended) \\
         \multicolumn{2}{l|}{Particle species} & Assign kernel size by particle mass, local density, or none\\
         Minimum number of particles per halo & $n_\mathrm{min}^\mathrm{halo}$ & Discard haloes below this number of particles (e.g., 25)\\
         \hline
         \multicolumn{3}{c}{Stellar haloes} \\ 
         \hline
         \multicolumn{2}{l|}{Component used for mesh halo finding} & Use only DM or DM+stars for identifying density peaks\\
         Kernel width for stars & $\ell_\mathrm{stars}$ & Interpolate stars in a cloud of radius $L/(N_x\times 2^{\ell_\mathrm{stars}})$ \\
         \multicolumn{2}{l|}{Minimum number of stellar particles per stellar halo} &  \\
         Density increase (from inner minimum) to cut the halo & $f_\mathrm{min}$ & See Sect. \ref{s:algorithm.stellar}\\
         Maximum radial distance without stars to cut the halo & $\ell_\mathrm{gap}$  & See Sect. \ref{s:algorithm.stellar} \\
         Minimum density (in units of $\rho_B(z)$) to cut the halo & $f_B$ & See Sect. \ref{s:algorithm.stellar}\\
         \hline
    \end{tabular}
    \label{tab:parameters}
\end{table*}

\subsection{Input data}
\label{s:algorithm.io}

Generally, the input data for \ASOHF consists on a list of DM particles, containing the 3-dimensional positions and velocities, masses and a unique, integer identifier of each of the $N$ particles. While \ASOHF was originally envisioned to be coupled to the outputs of \texttt{MASCLET} cosmological code \citep{Quilis_2004,Quilis_2020}, it can work as a fully stand-alone halo finder. The reading routine is fully modular and can be easily adapted to suite the users' input format\footnote{For more information, check the code documentation in \url{https://asohf.github.io}.}.

\subsection{Grid halo identification}
\label{s:algorithm.grid}
The identification of haloes relies on the analysis of the underlying continuous density field, which is obtained by means of a grid interpolation from the particle distribution. Originally inherited from \texttt{MASCLET}, since its original version \ASOHF uses an {\it Adaptive Mesh Refinement} (AMR) hierarchy of grids to compute the density field on different scales, thus enabling to capture a large dynamical range in masses and radii. In this section we describe the mesh creation procedure (Sect. \ref{s:algorithm.grid.creation}), which has been optimised to allow the code to handle even hundreds of thousands of refinement patches in large simulations; the new density interpolation scheme (Sect. \ref{s:algorithm.grid.density}) which mitigates the effect of sampling noise; and the halo finding process over the grid (Sect. \ref{s:algorithm.grid.haloes}). It is worth noting that, to avoid arbitrariness in the definition of a given structure, at this stage the new version of \ASOHF does not deal with peaks within haloes, which will be considered later on (see Sect. \ref{s:algorithm.substructure}).

\subsubsection{Mesh creation}
\label{s:algorithm.grid.creation}
A coarse grid of size $N_x \times N_y \times N_z$ covers the whole domain of the input simulation. On top of this base grid, an arbitrary number of mesh refinement levels is generated, by placing patches covering the regions with highest particle number density, each level halving the cell size with respect to the previous one. In particular, we flag as \textit{refinable} any cell hosting more than a minimum number of particles ($n_\mathrm{part}^\mathrm{refine}$). The mesh creation routine then examines all the refinable cells, from densest to least dense, and tries to extend the patch along each direction if the fraction of refinable cells amongst those added exceeds $f_\mathrm{refinable}^\mathrm{extend}$. Only patches with a given minimum size ($N_\mathrm{min}^\mathrm{patch}$) are accepted; otherwise, the region does not get refined. These quantities ($n_\mathrm{part}^\mathrm{refine}$, $f_\mathrm{refinable}^\mathrm{extend}$, and $N_\mathrm{min}^\mathrm{patch}$), as well as the number of refinement levels ($n_\ell$), are free parameters, which can be tuned to find a balance between memory usage and resolution. Generally, the latter can be fixed so that the peak resolution matches the force resolution of the simulation. We address the reader to Sec. \ref{s:mock_tests.scalability.npart} for a test showing how these parameters work with varying particle resolutions.

\subsubsection{Density interpolation}
\label{s:algorithm.grid.density}
In our experiments, we find that interpolating particles into cells using the standard Cloud-in-Cell (CIC) / Triangular-Shaped Cloud (TSC) schemes at the resolution of each AMR patch has a detrimental effect for the purpose of density peak finding, since it produces a very large amount of peaks due to shot noise. While these fake peaks would be gotten rid of in subsequent steps, when refining halo properties with particles (Sect. \ref{s:algorithm.particles}), they produce an overwhelming computational burden and load imbalance between different threads.

To avoid this, we compute the local density field by spreading each particle, using linear or quadratic kernels, in a cubic cloud with the same volume the particle sampled back in the initial conditions. That is to say, if different particles species (in terms of mass) are present, each one gets spread into the AMR grids according to their corresponding kernel sizes, i.e. the kernel radius $\Delta x_{\mathrm{kern},i}$ is set by 

\begin{equation}
    \Delta x_{\mathrm{kern},i}={L}/{2^{\lfloor \log_8 \frac{M_\mathrm{box}}{m_{p,i}}\rfloor}},
\end{equation}

\noindent with $L$ the box size, $M_\mathrm{box}$ the total mass in the box, and $m_{p,i}$ the mass of the particle $i$. Alternatively, for simulations with equal-mass particles, the kernel size can be chosen to be determined by the local density in the base grid cell occupied by each particle.

This procedure naturally produces a smooth density field, free of spurious noise while still capturing local features such as density peaks associated with a hierarchy of substructures.

\subsubsection{Halo finding}
\label{s:algorithm.grid.haloes}
Haloes are pre-identified as peaks, i.e. local maxima, in the density field, from the coarsest to the finest AMR levels. To do so, at each AMR level, \ASOHF lists all the (non-overlapping) cells simultaneously fulfilling:

\begin{itemize}
    \item The cell density is above the virial density contrast at a given redshift, $\Delta_\mathrm{m,vir}(z)$, computed according to the prescription of \cite{Bryan_Norman_98}. Incidentally, we also consider cells whose overdensity exceeds $\Delta_\mathrm{m,vir}(z)/6$, since the density interpolation could smooth the peaks in some situations. We have checked, with the tests in Sect. \ref{s:mock_tests.field}, that further lowering this value does not allow to recover any more haloes.
    \item The cell corresponds to a local maximum of the density field, i.e., it has the larger density among its $3^3$ neighbouring cells.
\end{itemize}  

The algorithm then iterates over them with decreasing overdensity. For each cell, the procedure can be summarised as:

\begin{enumerate}
    \item Check whether the cell is inside a previously identified halo. If it is, skip it; if not, continue (note that we do not look for substructure at this stage).
    \item Refine the location of the density peak using the information in higher (i.e., finer) AMR levels, if available. Check that the refined position does not overlap with previously identified haloes.
    \item Grow spheres of increasing radii until the enclosed overdensity, measured using the density field, falls below $\Delta_\mathrm{m,vir}$.
\end{enumerate}

This produces a first list of \textit{isolated} haloes (i.e., not being substructure), with an estimation of their positions, radii and masses, which will be refined further by using the whole particle list information in the following step. Note that, at this step, haloes can overlap, although no halo centre can be placed inside another halo. These peaks will be dealt within the substructure finding step (Sect. \ref{s:algorithm.substructure}).

\subsection{Halo refinement using particles}
\label{s:algorithm.particles}

After being identified, halo properties are refined using the particle distribution in several steps, described below. We refer the reader to Sect. \ref{s:mock_tests.field} for a test which shows the capabilities of \ASOHF to identify and recover the properties of haloes unbiased. While the basic process (collecting particles, unbinding and determination of the virial radius), common to most SO halo finders, was present in the original version, the new version of \ASOHF includes many refinements to improve the quality of the recovered catalogue (such as the recentring of the density peak or additional unbinding procedures), as well as other features to significantly decrease the computational cost and compute certain halo properties.

\paragraph{Recentring of the density peak.} Within the scheme described in Sect. \ref{s:algorithm.grid}, the density peak location, as identified within the grid, has an uncertainty in the order of the cell size of the finest grid covering the peak. In order to improve this determination, the first step consists on a refinement of the density peak position using particles.

To do so, we consider a cube, with centre on the estimated density peak, of side length $2 \Delta x_\ell$, with $\Delta x_\ell$ the cell size of the finest patch used to locate the density peak within the grid. We compute the density field in this volume using an ad-hoc $4^3$ cells grid, and pick as the corrected centre the position of the largest local maximum of the density field. The process is henceforward iterated, whilst more than 32 particles are found in the cube, each time halving the cube side length and centring it on the peak of the previous step. The position found by this procedure is not further changed in the process of halo finding.

\paragraph{Selection of particles.} In order to alleviate part of the computational burden associated with traversing the whole particle list, for each halo, only the particles in a sphere with mean density $\langle\rho\rangle = \min (\Delta_\mathrm{vir,m}(z), \, 200) \, \rho_B(z)$, with $\rho_B(z)$ the background matter density at redshift $z$, are kept. Subsequently, this list is sorted by increasing distance to the halo centre.

Additionally, to boost performance by decreasing the number of array accesses and distance calculations, which happen to be an important performance penalty as the number of particles in the simulation increases, particles are sorted according to their $x$-component before starting the whole process of halo finding. Then, only a small fraction of the particle list is traversed in order to select the halo particle candidates, in a way similar to a tree search algorithm in one dimension. Naturally, this effect becomes increasingly dramatic as the volume being analysed increases.

\paragraph{Unbinding.} A common feature of configuration-space halo finders is the necessity to perform a pruning of those particles which are dynamically unrelated to the haloes, since particles are initially collected using only spatial information. We perform two complementary, consecutive unbinding procedures:

\begin{itemize}
    \item \textit{Local escape velocity} unbinding. First, Poisson equation is solved in spherical symmetry for the particle distribution (see Appendix \ref{s.appendix.unbinding}), yielding the gravitational potential $\phi(r)$. The escape velocity is then estimated as $v_\mathrm{esc}(r) = \sqrt{-2 \phi (r)}$. Finally, any particle with velocity (relative to the halo centre of mass reference frame) exceeding its local escape velocity should be flagged as unbound and pruned from the halo.
    
    However, since the bulk velocity of the halo can be severely contaminated by the unbound component, we unbind particles in an iterative way, by pruning particles with speed relative to the centre of mass larger than $\beta v_\mathrm{esc}(r)$. In successive steps, we take $\beta=8,\,4$, and finally $\beta=\beta_\mathrm{final}\equiv 2$. Each time, the gravitational potential generated by --only-- bound particles is recomputed. The process is iterated with the last value of $\beta$ until no more particles are removed in an iteration. 
    
    We set $\beta_\mathrm{final}=2$, instead of getting rid of all particles whose speed exceeds the local escape value, since these \textit{marginally unbound} particles may get bound at a later step and will not drift away from the cluster immediately \citep[see, e.g., the discussion in][]{Knebe_2013}.
    
    \item \textit{Velocity space} unbinding. The standard deviation of the particles' velocities within the halo, $\sigma_v$, with respect to the centre-of-mass velocity, is computed. Particles whose 3-dimensional velocity differs by more than $\beta \sigma_v$ from the centre-of-mass velocity are pruned. Like in the previous scheme, $\beta$ is lowered progressively, from $\beta=6$ to $\beta = \beta_\mathrm{final} = 3$. In each step, the centre-of-mass velocity is recomputed for bound particles.
\end{itemize}

We refer the reader to Sect. \ref{s:mock_tests.unbinding} for two specific tests which show the complementarity of these two unbinding schemes.

\paragraph{Determination of spherical overdensity boundaries.} The spherical overdensity boundaries $\Delta_m\equiv \frac{\langle \rho \rangle}{\rho_\mathrm{B}(z)}=200,\,500,\,2500$, $\Delta_c\equiv \frac{\langle \rho \rangle}{\rho_\mathrm{crit}(z)}=200,\,500,\,2500$ (where $\rho_\mathrm{crit}(z)$ is the critical density for a flat Universe) and $\Delta_\mathrm{vir}$ \citep{Bryan_Norman_98} are precisely determined. The centre-of-mass velocity inside the refined $R_\mathrm{vir}$ boundary, as well as the maximum circular velocity, its radial position and the corresponding enclosed mass are also computed. Haloes with less than a user-specified minimum number of particles ($n_\mathrm{min}^\mathrm{halo}$) are regarded as \textit{poor haloes}, and are removed from the list.

\paragraph{Determination of halo properties.} Finally, several properties of the halo, inside $R_\mathrm{vir}$, are computed. These include:

\begin{itemize}
    \item Centre of mass position and velocity
    \item Velocity dispersion
    \item Kinetic energy
    \item Gravitational energy (either by direct sum or using a sampling estimate, see Appendix \ref{s.appendix.gravitational_energy}), and most-bound particle (which serves as a proxy for the location of the potential minimum)
    \item Specific angular momentum
    \item Inertia tensor and principal axes of the best-fitting ellipsoid
    \item Mass-weighted radial speed
    \item Enclosed mass profile
    \item List of bound particles, from which any other possible information can be directly computed.
\end{itemize}

\subsection{Substructure finding}
\label{s:algorithm.substructure}

Once all the non-substructure haloes have been identified, we proceed to search for substructures, which we define as systems centred on density peaks within previously identified haloes (either isolated haloes or substructures detected at a finer level of refinement). The process of substructure finding is nearly parallel to that of finding non-substructure haloes. Thus, here we shall only outline the differences with the aforementioned procedure. The most remarkable distinction relates to the choice of halo boundary. While non-substructure (i.e., isolated) haloes can be well-characterised by an enclosed density threshold, such a threshold may not exist if the halo is embedded in a larger-scale overdensity. While some finders use a change in the slope of the density profile to set the boundary of substructure \citep[e.g. AHF,][]{Knollmann_2009}, we find that such a rise in the slope is not always found and leads to arbitrarily large substructure radii. Instead, we use the \textit{Jacobi radius}, $R_\mathrm{J}$, defined in \cite{Binney_Tremaine} as the saddle point of the effective potential generated by the host-satellite system, as an estimation of the substructure extent (the region of space where the attraction towards the satellite overcomes the one of the host). We note this new scheme for substructure characterisation is integrally new to the revised version of the finder.

\paragraph{Search on the grid.} Cells above the virial density contrast are considered as candidates to substructure centres if they are a strictly defined local maximum (their density is larger than in any other of the $3^3-1$ neighbouring cells), they do not belong to a previously identified substructure at the same grid level, and they are not within $2\Delta x_\ell$ of their host halo centre, where $\Delta x_\ell$ is the grid cell size at the given refinement level. This last step is enforced to avoid arbitrarily detecting the same peak as a subhalo of itself, while allowing to recover more and more central substructures as long as there are refinement levels available covering the central region of the host.

For each centre candidate, after recentring, we choose as host the halo (at the highest hierarchy level\footnote{I.e., if a peak, candidate to correspond to a substructure, is inside a halo and inside a subhalo, the corresponding structure will be regarded as a subsubhalo, if finally accepted.}) which minimises the distance between host and substructure centres, $D$. We then compute $M$, the mass of the host within a sphere of radius $D$, by means of a cubic interpolation from the previously saved enclosed mass profiles, and get a rough estimation of the substructure extent by numerically solving Eq. \ref{eq:jacobi_approx},

\begin{equation}
    \left(\frac{R_\mathrm{J}}{D}\right)^3-\frac{m}{3M + m}=0,
    \label{eq:jacobi_approx}
\end{equation}

\noindent being $m$ the mass of the substructure candidate within a sphere of radius $R_\mathrm{J}$. This equation is a simplification of the exact definition of the Jacobi radius (see below, Eq. \ref{eq:jacobi_exact}; see also \citealp{Binney_Tremaine}) under the assumption $m \ll M$ (and $r_\mathrm{J} \ll D$). We note that an approximate version of this definition was already implemented by \texttt{MHF} \citep{Gill_2004}. 

\paragraph{Refinement with particles.} The procedure is analogous to the one described above, the sole difference being that the boundary of the halo (for measuring all halo properties) is taken as the Jacobi radius. In this respect, we use particles to refine this boundary with the non-approximate expression yielding $R_\mathrm{J}$, as given by \cite{Binney_Tremaine}:

\begin{equation}
    f(x) \equiv \frac{1}{(1-x)^2}-\frac{g(x)}{x^2}+\left[1+g(x)\right]x-1=0,
    \label{eq:jacobi_exact}
\end{equation}

\noindent with $x\equiv R_\mathrm{J}/D$ and $g(x) \equiv m/M$. After $R_\mathrm{J}$ has been identified, all halo properties can be computed from the bound particles as discussed in Sect. \ref{s:algorithm.particles}. We refer the reader to Sect. \ref{s:mock_tests.substructure} for a test of the substructure identification capabilities of \ASOHFns.

\subsection{Stellar haloes}
\label{s:algorithm.stellar}
Besides identifying DM haloes, the new version of \ASOHF is also able to characterise and produce catalogues of stellar haloes (i.e., galaxies) if such particles are provided. The identification of stellar haloes relies, in the first place, on the identification of the underlying DM haloes and subhaloes, which are typically more massive and less concentrated than their stellar counterpart \citep[e.g.,][]{Pillepich_2014}. However, it is worth emphasising that stellar haloes get then characterised by \ASOHF as independent objects. Thus, the procedure iterates over all previously found haloes and subhaloes, and performs the following steps.

\paragraph{Selection of particles. } All stellar particles inside the virial volume of the DM halo are collected, using the same tree-like search from the list of particles sorted along the $x$-coordinate as described in Sect. \ref{s:algorithm.particles}. All the bound DM particles identified in the halo finding step are recovered, and the whole list of DM+stellar particles is sorted by increasing distance to the centre of the DM halo. 

\paragraph{Determination of a preliminary boundary of the stellar halo. } In order to compute half-mass radii, we first need to place an outer boundary to the stellar halo. This is especially important in the case of central galaxies, where we do not want the masses, radii and other properties of the resulting galaxy to be affected by the presence of satellites or intracluster light. For each stellar halo candidate, we compute its spherically-averaged, stellar density profile, and place a radial cut at the smallest radius that fulfils one of the following conditions:

\begin{itemize}
    \item Stellar density increases by more than a factor $f_\mathrm{min}$ from the previous (inner) density minimum. This indicates the presence of a massive satellite.
    \item Stellar density falls below a given threshold, which we parametrise in terms of the background matter density as $f_B \rho_B(z)$.
    \item There is a gap in radial space, i.e., a comoving distance larger than $\ell_\mathrm{gap}$ without any stellar particle.
\end{itemize}

These three conditions are complementary, present small dependence on the free parameters and are conservative enough to avoid splitting a real stellar halo in several pieces. In our test, we find that the resulting galaxy catalogue is fairly independent of the density increase parameter, which can be varied in the range $f_\mathrm{min} \in [2,20]$. The results do not depend strongly on $f_\mathrm{B} \sim 1$, since stellar density profiles usually present a sharp boundary. However, we note that too low values ($f_\mathrm{B} \ll 1$) should be prevented, since they may add a large contamination by intracluster light stellar particles. Finally, $\ell_\mathrm{gap}$ can be set to a conservative value, $\sim (5-10) \, \mathrm{kpc}$.

\paragraph{Unbinding. } The unbinding steps are conceptually similar to the ones discussed in Sect. \ref{s:algorithm.particles} for DM haloes, but it is worth stressing a few subtleties. 

\begin{itemize}
    \item \textit{Local escape velocity} unbinding. The procedure is parallel to the one discussed for DM haloes. In this case, we consider all particles inside the previously found fiducial radius to solve Poisson's equation. However, we only unbind stellar particles, since all DM particles were already bound to the underlying DM halo by construction.
    \item \textit{Velocity space} unbinding. From this point, we get rid of DM particles and only consider the stellar ones. While this procedure is analogous to the one performed for the DM halo, by only considering stellar particles we account for the fact that DM and stellar components could correspond to distinct kinematic populations.
\end{itemize}

\paragraph{Determination of half-mass radius and recentring. } From the list of bound particles, the half-mass stellar radius is determined as the distance to the first particle whose enclosed bound mass exceeds half the total mass of bound particles inside the fiducial radius. Up to this point, the DM halo centre had been used as provisional centre. Henceforward, we perform an iterative recentring (analogous to the one described in Sect. \ref{s:algorithm.particles}) to the peak of stellar density, by iteratively looking at the largest stellar density peak inside the half-mass radius sphere. Finally, the half-mass radius is recomputed from this centre, which by definition yields a tighter radius than the first estimate. 

\paragraph{Characterisation of halo properties. } Finally, \ASOHF computes a series of properties of the galaxy. All these properties, which include the stellar-mass inertia tensor, stellar angular momentum, velocity dispersion of stellar particles, etc., are referred to the half-mass radius. The code can also output the whole list of stellar particles in the halo for further analyses.

We refer the reader to Sect. \ref{s:simulation_tests.stars} for results of the stellar halo finding capabilities of \ASOHFns.

\subsection{Additional features and tools}
\label{s:algorithm.additional}

Besides the main code, the \ASOHF package includes a series of complementary tools (mainly implemented in \texttt{python3} with the usage of standard libraries) to set-up a domain decomposition for running \ASOHF (Sect. \ref{s:algorithm.additional.domdecomp}), to compute merger trees from \ASOHF catalogues (Sect. \ref{s:algorithm.additional.mtree}), as well as a library to load all \ASOHF outputs into \texttt{Python}.

\subsubsection{Domain decomposition}
\label{s:algorithm.additional.domdecomp}

As we show in Sect. \ref{s:mock_tests.scalability.npart} below, the wall-time of our halo-finding code scales proportionally to the number of DM particles and haloes. Thus, especially when analysing large spatial volumes, performance can be greatly boosted by decomposing the domain. Each domain can be run by an independent \ASOHF process, and the resulting catalogues (together with any other output files) can be merged to get a single catalogue representing the whole input domain. The absence of communication between the different domains allows to run the different domains either sequentially in the same machine or concurrently in different machines, thus effectively allowing to run \ASOHF in distributed memory platforms.

To enable this procedure, \ASOHF allows the user to specify a rectangular subdomain as an input parameter. All particles outside this domain get discarded by the reader, so as to reduce memory usage. The \texttt{python} script \texttt{setup\_domdecomp.py}, included within the \ASOHF package, automates this task by creating the necessary folders, executables and parameter files for each domain. The number of divisions along each direction determines a \textit{fiducial domain} for each task. Each of these fiducial domains, which cover the whole domain without overlapping with each other, gets enlarged in all directions by an overlap length, which is also a free parameter. To avoid losing objects close to these boundaries of the domains, the overlap length can be safely set to the largest expected size of a halo in the simulation ($\sim 3 \, \mathrm{Mpc}$ for standard cosmologies). Once the domain decomposition is set up, we provide example shell scripts for running all the domains, either sequentially or concurrently (we provide an example \texttt{slurm} script).

Once the halo finding procedure has concluded for a given snapshot, the output files of each domain can be merged using the \texttt{merge\_domdecomp\_catalogues.py} script, which keeps all haloes whose centre lies on the fiducial domain. When substructure is present, the position of the progenitor halo (or the first non-substructure halo up the hierarchy) is considered to decide whether the substructure is kept in the merged catalogue. While the overlap between adjacent domains, if sufficiently large as discussed above, ensures that no structure is lost by the decomposition, the strategy of only keeping haloes whose density peak lies within the fiducial domain, or whose host fulfils these conditions, guarantees that no halo is identified twice.

\subsubsection{Merger trees}
\label{s:algorithm.additional.mtree}

Also included in the \ASOHF code package, the \texttt{mtree.py} \texttt{python3} script allows to build merger trees from the catalogues and particle lists files. For each pair of consecutive snapshots (hereon, the \textit{prev} and the \textit{post} iterations) of the simulations, for which the halo catalogues have already been produced, the script first identifies, for each \textit{post} halo, all the \textit{prev} haloes in a sphere with radius equivalent to the maximum comoving distance travelled by the fastest particle in either the \textit{prev} or the \textit{post} iteration. These will be referred to as the \textit{progenitor candidates}.

For each of the \textit{progenitor candidates}, the intersection of its member particles with the \textit{post} halo is computed using the unique IDs of the particles. This process is done in $\mathcal{O} (N_\mathrm{prev} + N_\mathrm{post})$ for each intersection, instead of $\mathcal{O}(N_\mathrm{prev} N_\mathrm{post})$, by having sorted the particle lists by ID at the moment of reading the catalogues, thus boosting the performance of the \texttt{mtree.py} code. For instance, for a simulation with $\sim 10^8$ particles and $\sim 20 \, 000$ haloes, connecting the haloes between each pair of snapshots takes $1-2$ minutes on 16 threads. 

We account as progenitors all haloes which have contributed more than a fraction $f_\mathrm{given}$ to the descendant mass, which we arbitrarily set to a sufficiently small value, such as $f_\mathrm{given}=10^{-3}$, for the merger tree to be complete. For each progenitor above this threshold, we report the following quantities:

\begin{itemize}
    \item Contribution to the descendant halo, $M_\mathrm{int}/M_\mathrm{post}$, where $M_\mathrm{int}$ is the mass of the particles both in the \textit{post} and the \textit{prev} halo. This quantity needs to be interpreted carefully for substructures, since we account the particles of a substructure also as belonging to its host. Therefore, most typically a substructure will be contributed close to $100\%$ by its host halo in the previous snapshot.
    \item Retained mass, $M_\mathrm{int}/M_\mathrm{prev}$, i.e. the fraction of \textit{prev} mass given to the descendant halo. Again, in the presence of substructures it is important to note that, in most situations, a host halo will be quoted as retaining $100\%$ of a substructure in the previous iteration.
    \item A third figure of merit is the normalised shared mass, $M_\mathrm{int}/\sqrt{M_\mathrm{prev} M_\mathrm{post}}$, which is the geometric mean of the two above quantities. This quantity has the virtue of supressing the links of very massive haloes with very small haloes and vice-versa, since they get either supressed by the largest of $M_\mathrm{prev}$ and $M_\mathrm{post}$.
\end{itemize}

Additionally, for tracking the main branch of the merger tree, we use the (approximate) determination of the most gravitationally-bound particle introduced in Sect. \ref{s:algorithm.particles} and described in Appendix \ref{s.appendix.gravitational_energy}. Thus, we check whether the most-bound particle of the \textit{post} halo is in the \textit{prev} candidate, and vice-versa. The two-fold check is necessary in the presence of substructure, since the most-bound particle of a substructure most usually lies within the host halo.

It may happen with some frequency, especially in the case of small haloes and substructures in very dense environments, that a halo is lost in one iteration, and recovered afterwards. In order to avoid losing the main branch of a DM halo due to this spurious effect, the \texttt{mtree.py} script allows to link these \textit{lost} haloes to their progenitors skipping an arbitrary number of iterations.

\subsubsection{Python readers}
\label{s:algorithm.additional.reader}
All the information contained in \ASOHF outputs can be easily loaded into \texttt{python} for analysis purposes, making use of the included \texttt{readers.py} library. Catalogue readers for, both, DM and stellar haloes files, which allow the user to read and structure the data in several useful formats. Particle lists can be as well loaded from \texttt{python}. These readers take care of the different indexing conventions between \texttt{Fortran} and \texttt{python}.


\section{Mock tests and scalability}
\label{s:mock_tests}
Before testing \ASOHF on actual simulation data and comparing its results to other halo finders, we have quantified the performance of the key procedures of the code in some idealised, yet rather complex, tests. In particular, we have focused on the identification of isolated haloes (Sect. \ref{s:mock_tests.field}), the identification of substructures (Sect. \ref{s:mock_tests.substructure}), the unbinding procedures (Sect. \ref{s:mock_tests.unbinding}) and the performance of the code (Sect. \ref{s:mock_tests.scalability}).

\subsection{Test 1. Isolated haloes}
\label{s:mock_tests.field}
The first idealised test aims to prove the ability of the code to identify haloes in a broad range of masses, without overlaps nor substructure. The setup of the test is as follows.

We consider a flat $\Lambda$CDM cosmology, with $\Omega_\mathrm{m}=0.31$, $\Omega_\Lambda=0.69$, $h\equiv H/(100\unit{km \, s^{-1} \, Mpc^{-1}})=0.678$ and $\sigma_8=0.82$, at redshift $z=0$, and a cubic domain of side length $L=40\unit{Mpc}$. This domain will be populated with $N_\mathrm{part}$ particles of equal mass, $m_\mathrm{p}=\rho_\mathrm{B}(z=0)L^3 / N_\mathrm{part}$. Halo masses are drawn from a \cite{Tinker_2008} mass function by inverse transform sampling, setting a lower mass limit of $M_\mathrm{min}=50 m_\mathrm{p}$. The sampling is constrained so as to produce one halo with mass above $8 \times 10^{14} M_\odot$. The number of haloes is set by integrating the mass function from $M_\mathrm{min}$. Their corresponding virial radii are computed and haloes are placed at random positions, avoiding overlaps and crossing the box boundaries. Each halo is then realised with particles, by sampling a NFW profile \citep{NFW} for the radial coordinate, using the concentration-mass ($c_\mathrm{vir}-M_\mathrm{vir}$) relation modelled by \cite{Ishiyama_2021}, and assuming spherical symmetry for the angular coordinates. The NFW profiles are extended up to $1.5 R_\mathrm{vir}$ to avoid a sharp cut of the density profile, and the remaining particles up to $N_\mathrm{part}$, after sampling all haloes, are placed at random positions outside them. For this test, we use $N_\mathrm{part}=128^3$, so that 1528 haloes above $M_\mathrm{min} \approx 6 \times 10^{10}\, M_\odot$ are generated and realised with particles of mass $m_p=1.2\times 10^{9} \, M_\odot$. These parameters are varied in Sect. \ref{s:mock_tests.scalability}, when considering the scalability of the code.

Note that, while complex due to the high number of haloes, this test is idealised in the sense that it lacks any large-scale structure (LSS), such as filaments connecting haloes. As a matter of fact, this limitation of the test design makes it more challenging to detect small, isolated haloes, since they may only occupy one base grid cell and not get refined enough to be detected as a density peak\footnote{The situation is different for a realistic simulation output, where the presence of a web of filaments surrounding a low-mass halo may more easily trigger the creation of a refinement patch covering it.}. Therefore, the key parameter in this test is the base grid resolution, $N_x$. The rest of parameters have been fixed to $n_\ell=4$, $n_\mathrm{part}^\mathrm{refine}=3$, and $N_\mathrm{min}^\mathrm{patch}=14$ and have very limited impact on the outcomes of the test.

\begin{figure}
    \centering
    \includegraphics[width=\linewidth]{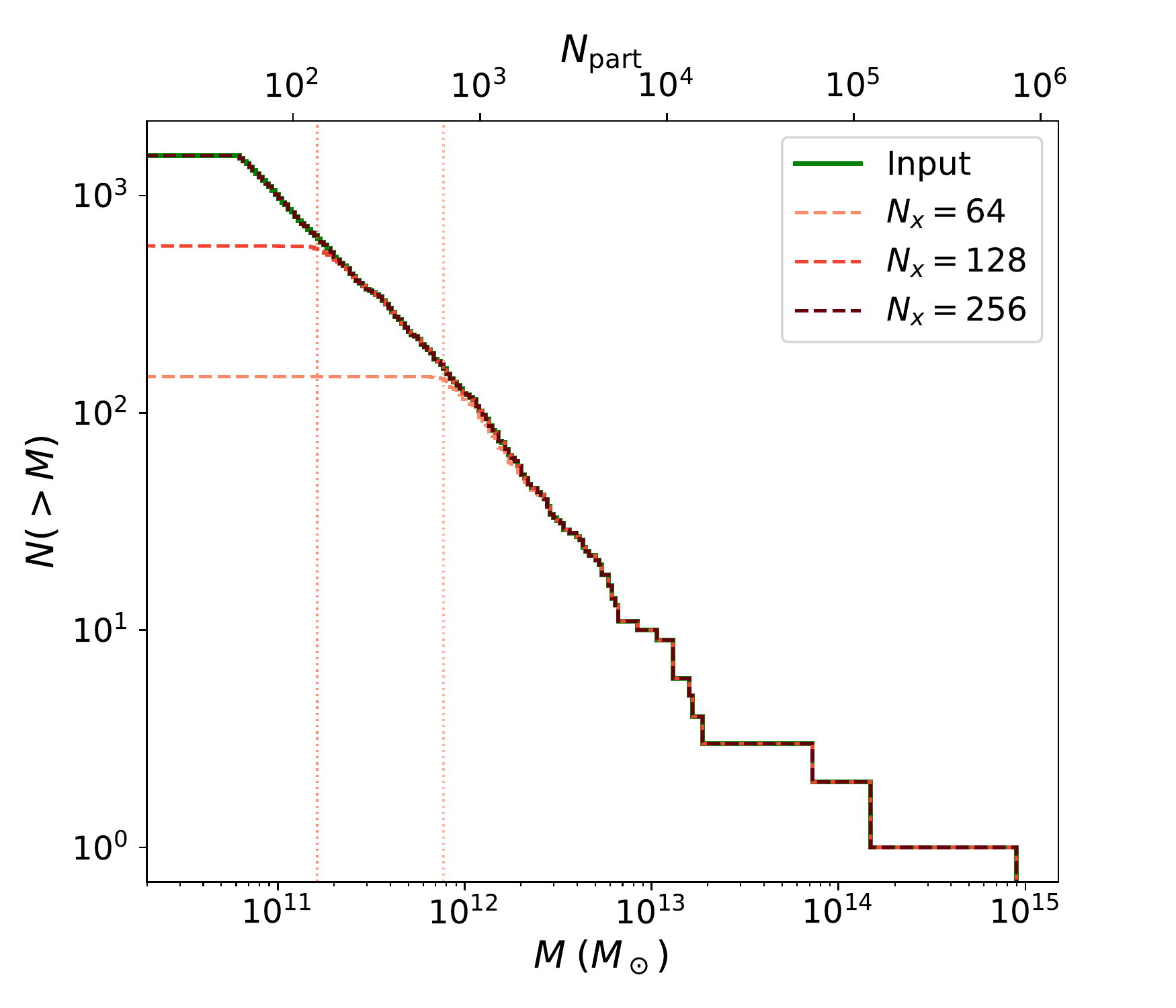}
    \caption{Results from \textit{Test 1}. Input cumulative mass function (green line) compared to \ASOHF results with $N_\mathrm{x}=64$ (light red), $N_\mathrm{x}=128$ (red) and $N_\mathrm{x}=256$ (dark red). The vertical lines mark the completeness limit, at $90\%$, of the catalogue produced by \ASOHFns. This value is not reported with $N_\mathrm{x}=256$, since the code is able to detect all haloes.}
    \label{fig1}
\end{figure}

\begin{figure*}
    \centering
    \includegraphics[width=\textwidth]{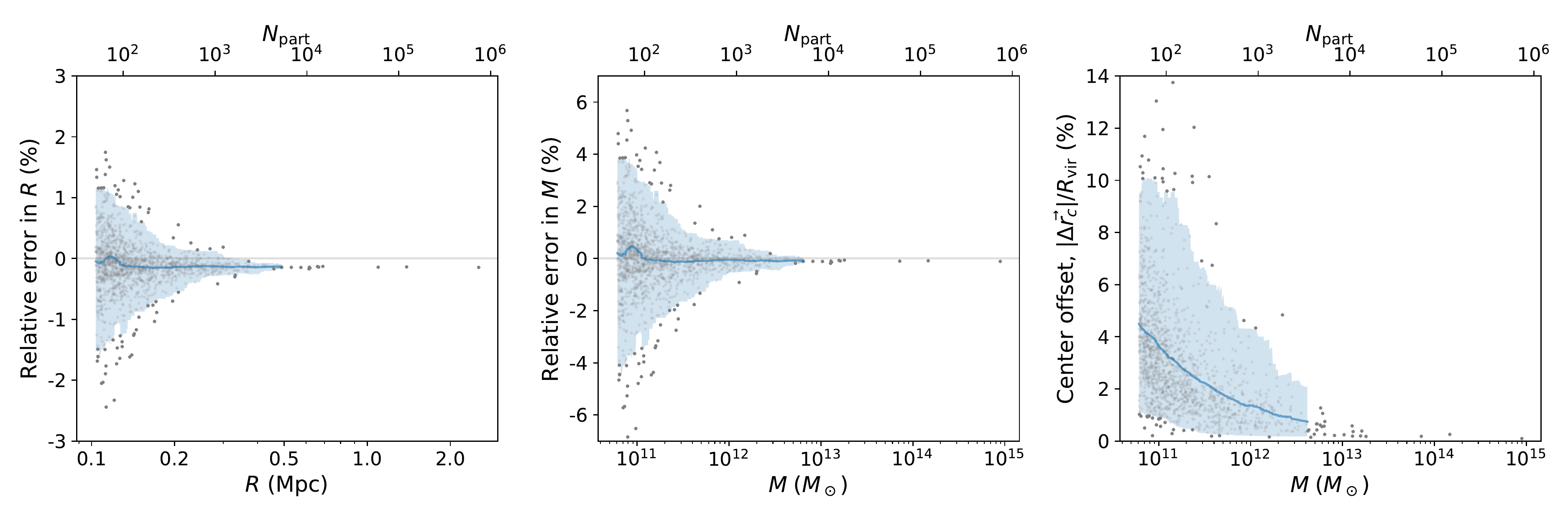}
    \caption{Precision of \ASOHF in recovering basic halo properties in \textit{Test 1}, with $N_x=256$. In each panel, dots represent individual haloes, the blue line presents the smoothed trend by using a moving median, and the shaded region encloses the $2\sigma$ confidence interval around it. \textit{Left panel:} relative error in the determination of the virial radius. \textit{Middle panel:} relative error in the determination of the virial mass. \textit{Right panel:} centre offset, in units of the virial radius.}
    \label{fig2}
\end{figure*}

Fig. \ref{fig1} presents the unnormalised mass functions (number of haloes with mass greater than $M$, $N(>M)$) of the catalogues generated by \ASOHF for $N_\mathrm{x}=64,\,128,\,\text{and}\,256$ (in light red, red and dark red dashed lines, respectively), compared to the input (thick, green line). Due to the overabundance of small haloes, the number of haloes detected is not an easily interpretable measure of the performance of the algorithm. Instead, we quote the \textit{completeness limit} of the sample detected by \ASOHFns, defined as the largest mass $M_\mathrm{lim}$ so that the recovered mass functions differs more than some fraction $1-\alpha$ from the input one. This results are listed, at completeness $\alpha=0.9$, in Table \ref{tab:completeness_1}, and represented as vertical, dotted lines in Fig. \ref{fig1}. With increasing resolution, the algorithm is capable of systematically  detecting less-massive haloes, being able to detect all 1528 of them when using $N_x=256$.

\begin{table}
    \centering
    \caption{Completeness limits of \ASOHF halo finding for \textit{Test 1} (in mass and number of particles, $M_\mathrm{lim}$ and $N_\mathrm{part}^\mathrm{lim}$), at $90\%$. With $N_\mathrm{x}=256$, all haloes are detected and so we do not report these limits.}
    \begin{tabular}{c|c|c}
         $N_\mathrm{x}$ & $M_\mathrm{lim}\; (M_\odot)$ & $N_\mathrm{part}^\mathrm{lim}$  \\ \hline
         64 & $7.71 \times 10^{11}$ & 638 \\
         128 & $1.63 \times 10^{11}$ & 134 \\
         256 & All & All
    \end{tabular}
    \label{tab:completeness_1}
\end{table}

The precision of such identification is shown in Fig. \ref{fig2}, where we have matched the input and output catalogues (for the $N_x=256$ case) and test the ability of \ASOHF obtaining a precise estimation of radii (left panel), masses (middle panel) and halo centres (right panel). The left and central panels show that, on average, we get an unbiased estimation of radii and masses, although the scatter becomes larger when there are less than $\sim 1000$ particles, reaching an amplitude of $\sim 1.5\%$ for radius and $\sim 4.5\%$ for mass at the $95\%$ confidence level for haloes of 50--100 particles. This is mostly associated with the uncertainty in determining the halo centre, as shown in the right panel of Fig. \ref{fig2}. For haloes below $\sim 1000$ particles, the median offset between the input and the recovered density peak differs by $2\%$ of the virial radius, increasing up to $\lesssim 10\%$ at the 97.5-percentile error for haloes with $\sim 50$ particles. This effect is expected, since the density in a sphere of given comoving radius decreases with decreasing mass for NFW haloes. Related to this, there is an unavoidable source of uncertainty in the test set-up, for low-mass haloes, due to the fact that we are sampling the particles from a probability distribution when creating the test (thus, the nominal centre may differ from the centre of the realisation with particles).

\subsection{Test 2. Substructure}
\label{s:mock_tests.substructure}
To assess the ability of the code to detect substructure, we have designed a second test consisting on a large, massive halo rich in substructure, with a similar procedure as in the previous test. In particular, we consider the same domain and cosmology, and place a halo with mass $10^{15} M_\odot$ in its centre. In order to being able to span a wide range in substructure masses, we use ${N_\mathrm{part}=512^3}$ DM particles, each with mass $m_\mathrm{p}=1.9 \times 10^7 M_\odot$. The particles in this host halo are generated in the same way as in Test 1, up to $1.5 R_\mathrm{vir}$.

Subsequently, we place $N_\mathrm{subs}=2000$ substructures, with masses drawn from the same \cite{Tinker_2008} mass function, from $M_\mathrm{min}=50 m_p = 9.4 \times 10^{8} \, M_\odot$ and constraining the sample to have at least one large subhalo, with mass above $10^{13} \, M_\odot$. Subhaloes are placed uniformly inside the host volume, avoiding overlaps, and are populated with particles following the same procedure as we have done with isolated haloes. These particles are superimposed on top of the particle distribution of the host. The remaining particles up to $N_\mathrm{part}$ are placed outside the host halo by sampling a uniform, random distribution to constitute a homogeneous background. It is worth stressing that, even though the particles of each halo are sampled from a spherically-symmetric NFW profile, the resulting realisation of the halo can depart strongly from spherical symmetry due to sample variance (especially relevant in smaller haloes, which are the most abundant). Therefore, the test contains many non-spherical haloes, including elongated systems which could resemble tidally stripped haloes.

\begin{figure}
    \centering
    \includegraphics[width=\linewidth]{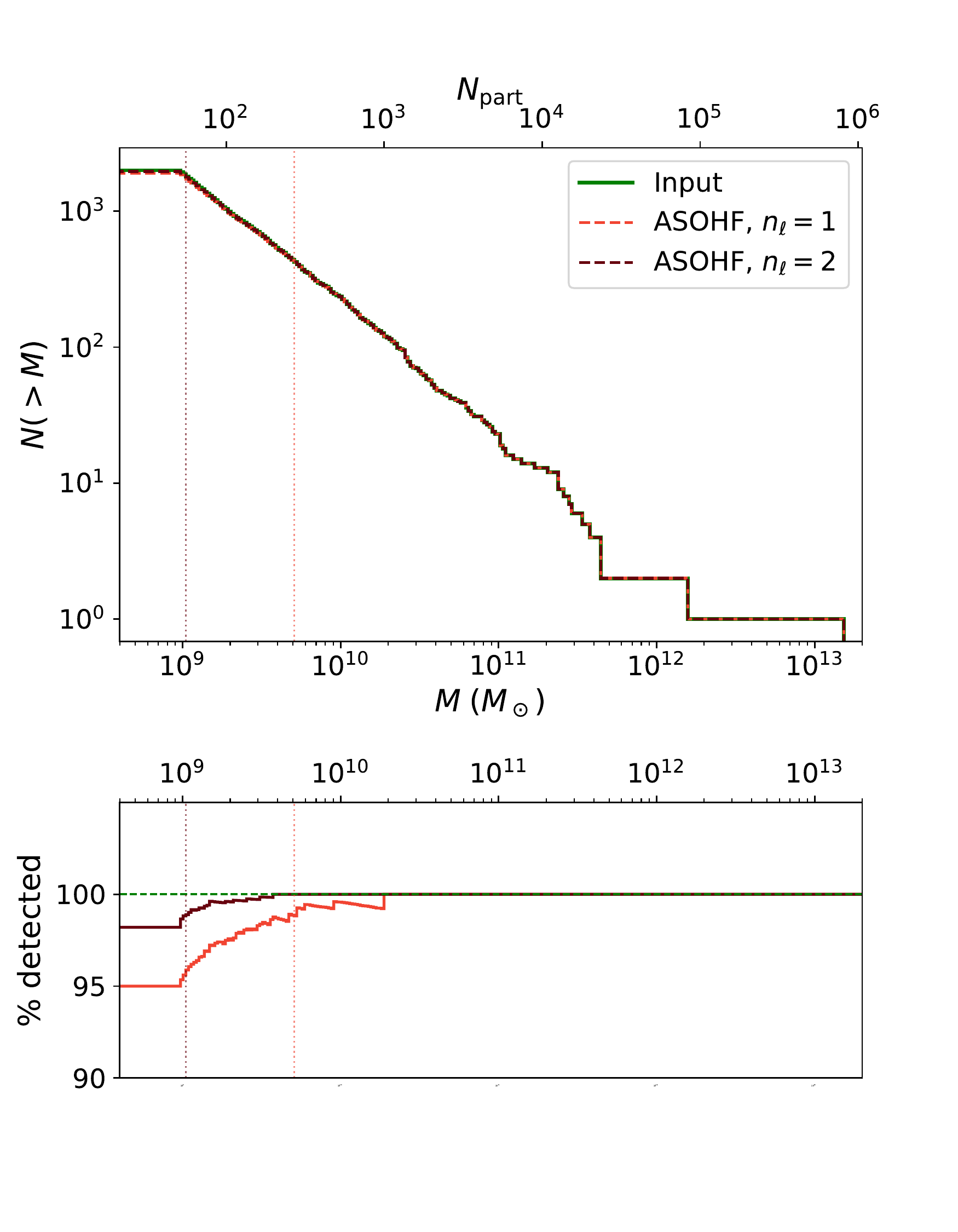}
    \caption{Results from \textit{Test 2}. \textit{Upper panel:} Input substructure cumulative mass function (green line) compared to \ASOHF results with $n_\ell=1$ (red) and $n_\ell=2$ (dark red). The vertical lines mark the completeness limit, this time at $99\%$, of the catalogue produced by \ASOHFns. \textit{Bottom panel:} Fraction of substructures with input mass larger than $M$ detected by \ASOHFns, with the same colour codes as above.}
    \label{fig3}
\end{figure}

We have run \ASOHF on this particle distribution using a base grid of $N_\mathrm{x}=512$ cells in each direction and $n_\ell=1 \text{ and } 2$ refinement levels, with a threshold of $n_\mathrm{min}^\mathrm{halo}=25$ particles per cell to flag it as refinable and accepting all patches with at least $N_\mathrm{min}^\mathrm{patch}=14$ cells in each direction. The results, in terms of detection capabilities of \ASOHFns, are shown in Fig. \ref{fig3}. Since the input and recovered mass functions visually overlap, the results are better assessed looking at the bottom panel, which shows their quotient. In general terms, using just one refinement level allows to detect $95\%$ of substructures, while using two levels for the mesh increases this fraction to over $98\%$. In terms of completeness limits, this time taking a more restrictive value of $\alpha=0.99$, these correspond to $\sim 270$ and $55$ particles, respectively, for $n_\ell=1,\, \text{and} \, 2$. The fact that these limits, even at a more restrictive threshold, are better than those in Test 1 (Table \ref{tab:completeness_1}) is due to the dense environment the substructures are embedded into, which more easily triggers the refinement of these regions.

\begin{figure}
    \centering
    \includegraphics[width=1.05\linewidth]{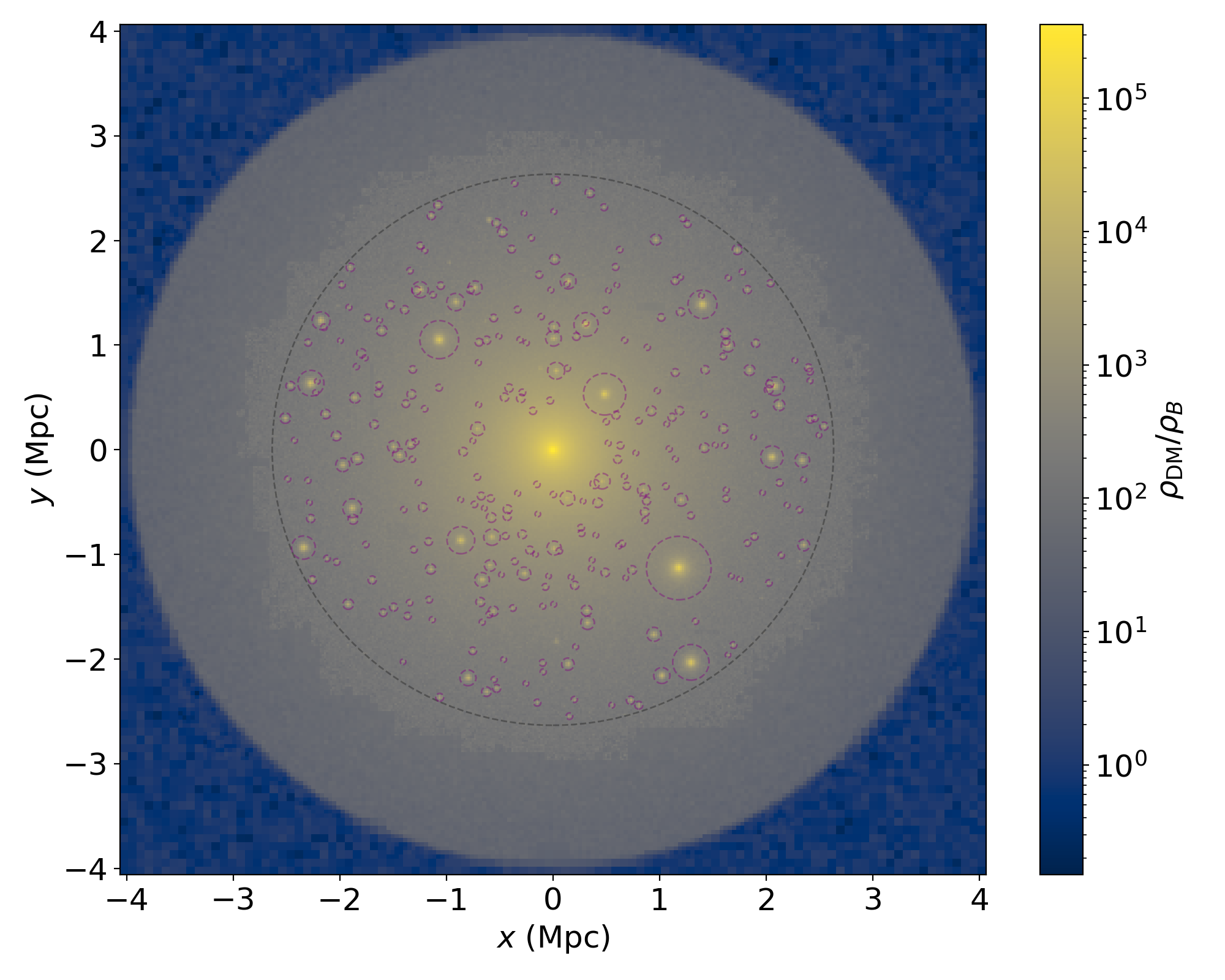}
    \caption{Projection, $460 \, \mathrm{kpc}$ thick, of the DM density field interpolated by \ASOHFns\ in \textit{Test 2}, from which the pre-identification of haloes over the AMR grids is performed. The plot shows the maximum value along the projection direction. The black circle marks the virial radius of the host halo, while each purple circle corresponds to a substructure identified by \ASOHFns, the radius of the circle matching the Jacobi radius of the subhalo.}
    \label{fig4}
\end{figure}

\begin{figure}
    \centering
    \includegraphics[width=\linewidth]{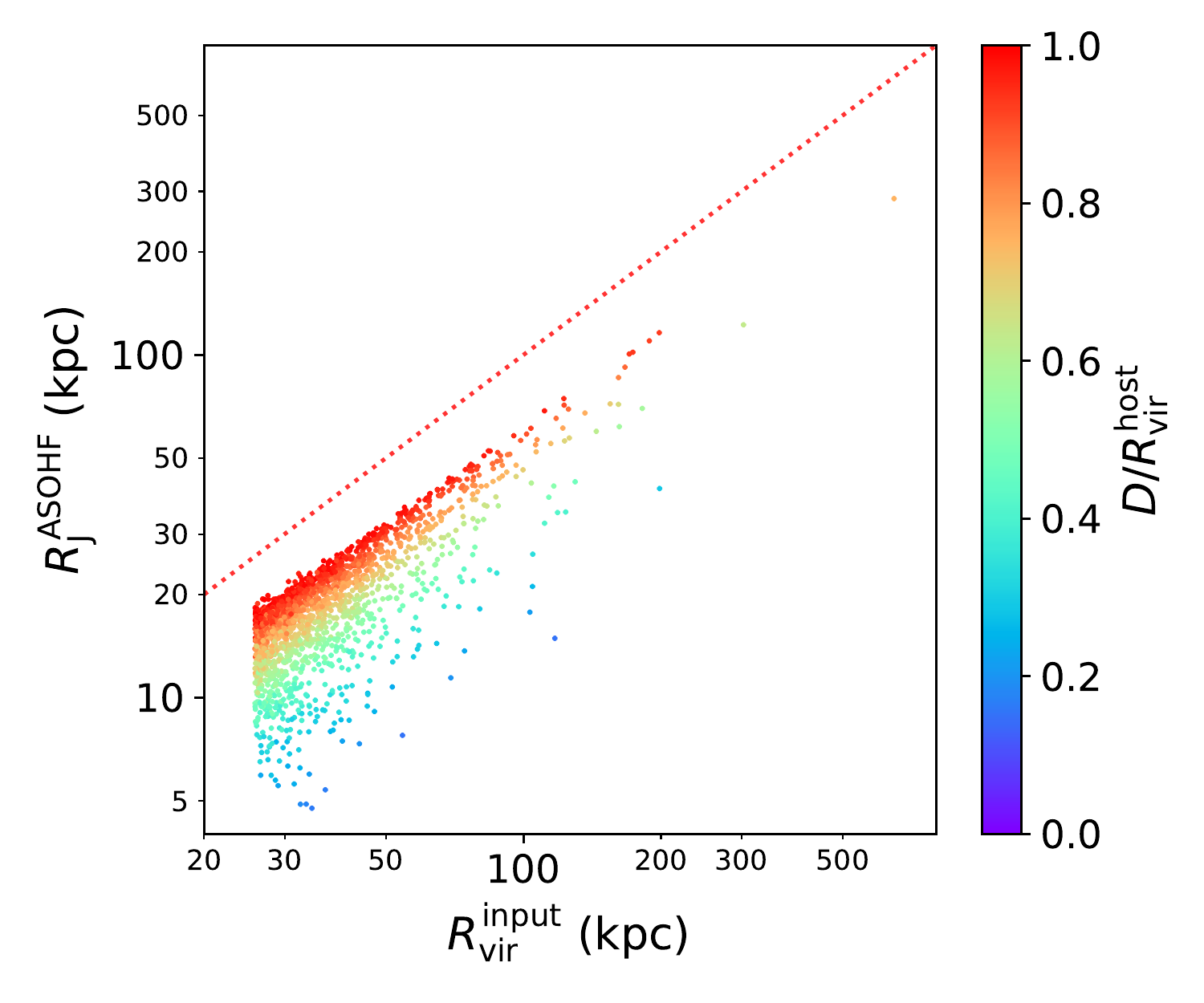}
    \includegraphics[width=\linewidth]{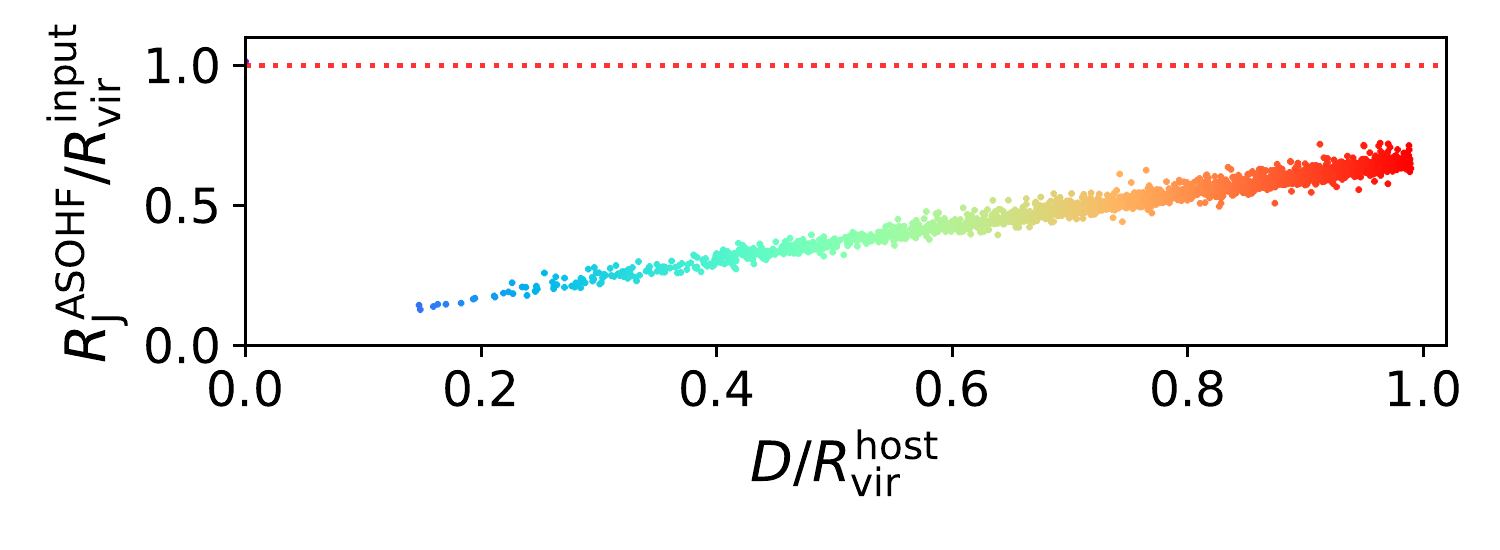}
    \caption{Relation between virial and Jacobi radii in \textit{Test 2}. \textit{Top panel:} Comparison between the virial radius of the input NFW halo ($R_\mathrm{vir}^\mathrm{input}$) and the Jacobi radius recovered by \ASOHF ($R_\mathrm{J}^\mathrm{ASOHF}$) for all substructures detected. Colours encode the radial position of the substructure centre in the host halo ($D$) in units of the host virial radius ($R_\mathrm{vir}^\mathrm{host}$). The dotted, red line corresponds to the identity relation, $R_\mathrm{J}^\mathrm{ASOHF}=R_\mathrm{vir}^\mathrm{input}$. \textit{Bottom panel:} Closer look at the tight linear relation between the quotient $R_\mathrm{J}^\mathrm{ASOHF}/R_\mathrm{vir}^\mathrm{input}$ and the radial position of the substructure.}
    \label{fig5}
\end{figure}

Fig. \ref{fig4} presents a thin ($\sim 460 \, \mathrm{kpc}$) slice of the density field as computed by \ASOHF in order to pre-identify haloes. The adaptiveness allows to simultaneously capture small substructures within the dense host halo, while getting rid of sampling noise in underdense regions, which would increase and unbalance the computational cost. Purple circles represent the extent of the substructure, i.e., a sphere of radius $R_\mathrm{J}$. We note that, in contrast to the virial radius of isolated haloes, the Jacobi radius naturally depends on the location of the substructure within its host: the more central its position, the smaller $R_\mathrm{J}$ will be in relation to the input virial radius of the NFW halo. This is quantitatively shown in Fig. \ref{fig5}, whose upper panel presents the relation between the input virial radius and the Jacobi radius determined by \ASOHFns, and implies that a satellite moving through a host would present a pseudo-evolution of $R_\mathrm{J}$ (and its enclosed mass) even if moving rigidly through the medium (the smaller the distance to the host, the smaller $R_\mathrm{J}$ is). While approximate due to the spherical symmetry and Keplerian rotation assumptions, this definition for the boundary of a substructure is reasonable since, during the dynamical evolution of the infall of a satellite, it is expected that the particles in the outer layers get unbound due to dynamical friction with the particles of the host. For more detailed studies, the whole list of particles of the satellite before its infall, which is also outputted by \ASOHFns, can be tracked (similarly to, e.g., \citealp{Tormen_2004}). In the lower panel, the tight linear correlation between the ratio of radii, $R_\mathrm{J}/R_\mathrm{vir}$, and the distance to the host centre, $D/R_\mathrm{vir}^\mathrm{host}$, is shown. This can be parametrised roughly as

\begin{equation}
    \frac{R_\mathrm{J}}{R_\mathrm{vir}} = (0.073\pm0.002) + (0.592\pm 0.003) \frac{D}{R_\mathrm{vir}^\mathrm{host}},
\end{equation}

\noindent although the parameters obviously dependend on the particular density profile of the host, and the scatter would be increased by the asphericity of real haloes.

\subsection{Test 3. Unbinding}
\label{s:mock_tests.unbinding}
The unbinding procedure is a critical step in all configuration space based halo finders, since the initial assignment of particles has neglected any dynamical information. Here we present two tests to prove the capabilities of the two complementary unbinding procedures implemented in \ASOHFns: in Sect. \ref{s:mock_tests.unbinding.a} we study the case of a small halo moving in a dense medium, while in Sect. \ref{s:mock_tests.unbinding.b} a fast stream traversing a halo is considered.

\subsubsection{Test 3a. Halo moving in a dense medium in rest}
\label{s:mock_tests.unbinding.a}
We consider a set-up similar to \textit{Test 1} (Sect. \ref{s:mock_tests.field}), in terms of box size and number of particles. We place a halo of virial mass $M_\mathrm{h}=5\times 10^{13} \, M_\odot$ in the centre of the box, and realise it with particles up to a radial distance of $3 R_\mathrm{vir} \approx 2.9 \, \mathrm{Mpc}$. This involves $73\, 976$ particles, which will hereon be referred to as \textit{halo particles}; $41\,428$ of which are inside the virial radius.

Then, the remaining $\sim 2 \times 10^6$ particles, corresponding to most of the mass in the box, are placed in uniformly random positions in a sphere of radius $6 R_\mathrm{vir}$ and will be referred to as \textit{background particles}. This amounts for a background of constant density $\sim 75 \rho_\mathrm{B}$, so that the halo and background densities are approximately equal at the virial radius of the cluster. Therefore, roughly $1/5$ of the particles inside the virial volume are background particles, which may bias many of the halo properties (centre of mass position, bulk velocity, angular momentum, etc.).

We have given the halo particles a bulk velocity of $3\,000 \, \mathrm{km\, s^{-1}}$ along the $x$-axis, so that background particles should be clearly unbound to the halo; while all particles (halo and background) are given a normal velocity dispersion with standard deviation of $300 \, \mathrm{km\, s^{-1}}$ to add some noise. We note that it is not the aim of this test to use physically realistic values, but just to show the robustness and performance of the unbinding procedure in a fairly reasonable situation.

\begin{figure*}
    \centering
    \includegraphics[width=\textwidth]{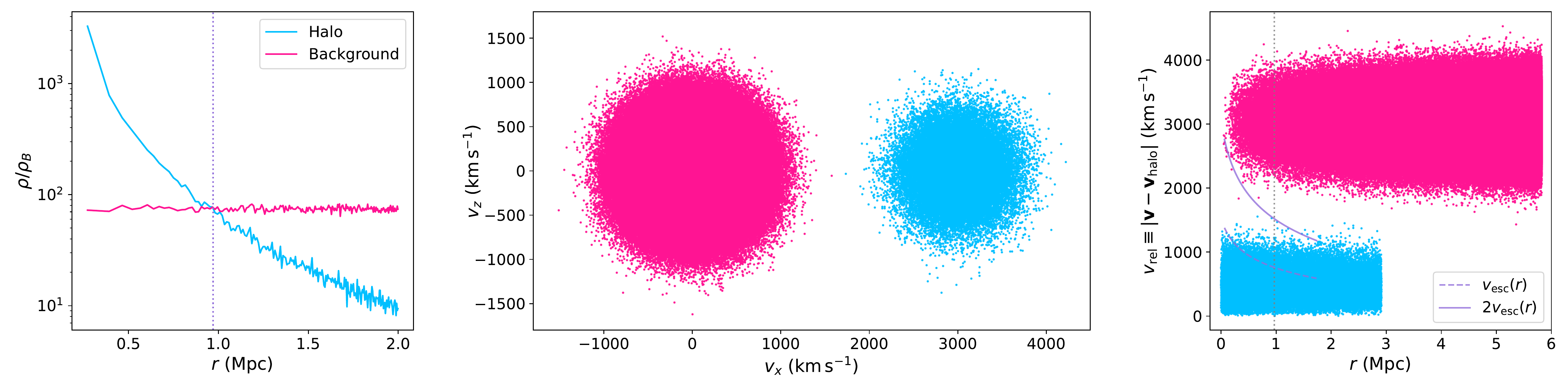}
    \caption{Results from \textit{Test 3a} (unbinding). \textit{Left panel:} Density profile of halo particles (blue) and background particles (pink). The dashed, vertical line marks the input virial radius of the halo. \textit{Middle panel:} $v_z - v_x$ phase space, using the same colour coding as in the previous panel. The particle distributions are disjoint in velocity space. \textit{Right panel:} $v_\mathrm{rel}-r$ phase plot, where the unbinding is performed. The same colour coding as in the previous panels is used, the vertical line marks the input virial radius, and the dashed and solid purple lines are $v_\mathrm{esc}(r)$ and $2 v_\mathrm{esc}(r)$, at the last iteration of the local velocity speed unbinding, the latter being the threshold velocity for unbinding.}
    \label{fig6}
\end{figure*}

Fig. \ref{fig6} summarises the set-up and the results of the test. Its left panel shows the density profiles of the halo particles (blue) and of background particles (pink), which roughly agree with each other at the virial radius of the halo. The central panel in Fig. \ref{fig6} presents the $v_z - v_x$ phase space, with the same colour code, and shows how these two components are totally disentangled in velocity space, while they are mixed up in configuration space. Finally, the right panel shows the $v_\mathrm{rel}-r$ phase plot, where $v_\mathrm{rel}\equiv |\vec{v}-\vec{v_\mathrm{halo}}|$, which corresponds to the space where the escape velocity unbinding is performed. In this plot, the dashed purple line represents the local escape velocity at the last iterative unbinding step (when no more particles have been unbound), while the solid purple line is twice this value (which is the last threshold speed used for the escape velocity unbinding, according to the procedure described in Sect. \ref{s:algorithm.particles}). Specifically, the particle-wise results are as follows:

\begin{itemize}
    \item The offset between the input centre and the one detected by \ASOHF is $|\Delta \vec{r}| = 2.3 \, \mathrm{kpc}$, which is $2.3 \permil$ of the virial radius (or $1.5\%$ of the scale radius of the NFW profile). This small miscentrering causes a $1.3\permil$ decrease in the host virial radius. Two halo particles which were nominally outside the halo to be listed as inside, while 29 inside particles appear to be outside.
    \item All background particles get pruned by one of the unbinding methods.
    \begin{itemize}
        \item Almost all background particles inside the virial volume (all but six) are pruned by the local escape velocity method, as shown by the fact that they lie above the threshold value $2 v_\mathrm{esc}(r)$ (solid purple line) in the right panel of Fig. \ref{fig6}. It is worth noting that the iterative procedure of lowering the threshold, as described in Sect. \ref{s:algorithm}, is crucial to prevent a biased centre-of-mass velocity in the first unbinding steps.
        \item The remaining six particles get pruned by the non-local unbinding in velocity space, since they lie at more than $3 \sigma_v$ of the centre-of-mass velocity. This method is especially useful for unbound components at small halo-centric distances, since escape velocities are high near the halo core.
    \end{itemize}
    \item Only one halo particle gets incorrectly pruned, for being slightly over $3 \sigma_v$ of the mean velocity.
\end{itemize}

This shows the ability of \ASOHF of pruning the unbound component of haloes, even when it represents a significant amount of the mass within the halo volume,and also illustrates the situation of a satellite moving through its host. We note that a crucial step, in order to being able to unbind host particles using the iterative procedure described in Sect. \ref{s:algorithm.particles}, is that the mass density of halo particles within the radial extent of the halo/satellite is greater than that of background or host particles. Otherwise, the centre-of-mass velocity would converge to that of the background. However, in the case of substructures this condition is automatically guaranteed by the definition of the Jacobi radius (note that Eq. \ref{eq:jacobi_approx} implies that the substructure is at least 3 times denser than the mean density of the host within a sphere of radius $D$).

\subsubsection{Test 3b. Fast stream traversing a halo}
\label{s:mock_tests.unbinding.b}
\begin{figure*}
    \centering
    \includegraphics[width=\textwidth]{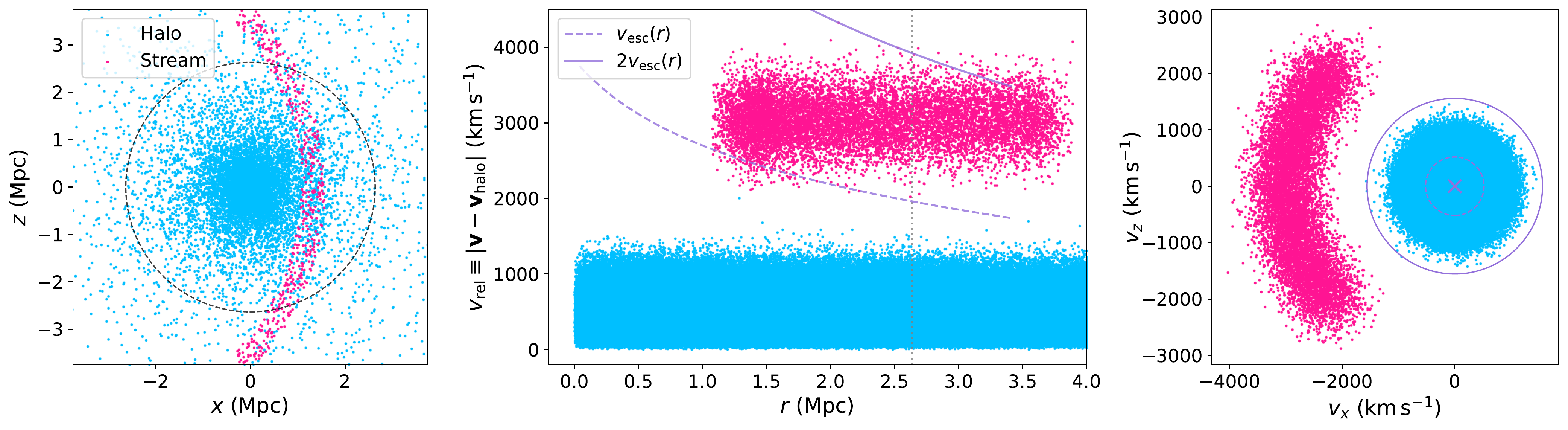}
    \caption{Results from \textit{Test 3b} (unbinding). \textit{Left panel:} Particles in a 20 kpc slice through the centre of the halo. Blue (pink) dots refer to halo (stream) particles. The gray, dashed circle corresponds to the location of the input virial radius. \textit{Middle panel:} $v_\mathrm{rel}-r$ phase plot, with the same colour coding as in the previous panel, showing that the local escape velocity unbinding does not succeed in pruning the stream. \textit{Right panel:} $v_z-v_x$ phase space, keeping the colour coding as in the previous plots. The purple cross, dashed circle and solid circle indicate, respectively, the converged centre-of-mass velocity, and the $1 \sigma_v$ and $3 \sigma_v$ regions. Particles outside the latter are pruned by the \textit{velocity space} unbinding.}
    \label{fig7}
\end{figure*}

To illustrate the importance of the complementary velocity space unbinding, we present here a second test, in which a fast stream traverses a halo at rest. The set-up is as follows. A halo of virial mass $M_\mathrm{vir}=10^{15} \, M_\odot$ is placed in the centre of a box, using the same particle mass ($m_\mathrm{p} \approx 1.2 \times 10^9 \, M_\odot$) as in the previous test. The halo is realised up to $3 R_\mathrm{vir}$ with $\sim 1.5 \times 10^6$ particles (the \textit{halo particles}). As for the stream of particles, we have considered a curved, cylindrical stream, with impact parameter $b=0.5 R_\mathrm{vir}$, curvature radius $r=2 R_\mathrm{vir}$, radius $\Delta b = 250 \, \mathrm{kpc}$ and length $L \approx 8.3 \, \mathrm{Mpc}$ (so that it crosses the whole virial volume of the halo). We assign a density of $200 \rho_\mathrm{B}$ to the stream, so that it amounts for a mass of $1.29 \times 10^{13} \, M_\odot$, or slightly over $10\,000$ particles (the \textit{stream particles}). The situation in configuration space is depicted in the left panel of Fig. \ref{fig7}, where blue (pink) dots represent the $x-z$ positions of halo particles (stream particles) lying within a $20 \, \mathrm{kpc}$ slice passing through the centre of the halo.

Velocities for the halo particles are drawn from a normal distribution with zero mean and an isotropic $\sigma_{1\mathrm{D}}= 300 \,\mathrm{km \, s^{-1}}$. Stream particles are given a speed of $3000 \, \mathrm{km \, s^{-1}}$ along the axis of the stream, plus an isotropic dispersion component as in the halo particles. It can be seen, in the middle panel of Fig. \ref{fig7}, that while most of the stream particles are nominally unbound, ($v > v_\mathrm{esc}(r)$), they do not reach the threshold for unbinding, $2 v_\mathrm{esc}$. While lowering this threshold would get rid of these particles, \textit{loosely unbound} particles (i.e., those with $v_\mathrm{esc}(r) < v < 2 v_\mathrm{esc}(r)$) may still remain within the halo for a dynamical time, making lowering the threshold somewhat aggressive (see also the discussion in \citealp{Knebe_2013} and \citealp{Elahi_2019}).

However, the situation in velocity space clearly presents two disjoint components, as represented in the right panel of Fig. \ref{fig7}. In this case, the \textit{velocity space} unbinding is able to get rid of all stream particles, since they lie beyond $3 \sigma_v$ of the centre-of-mass velocity after the iterative procedure.

\subsection{Scalability}
\label{s:mock_tests.scalability}

While the previous set of tests demonstrate the ability of \ASOHF in providing complete samples of haloes with unbiased properties, here we take a closer look at the performance of the code, in terms of execution time and memory requirements. We have used a set-up similar to that of Test 1 to assess how the code scales with the number of particles and base grid size (Sect. \ref{s:mock_tests.scalability.npart}), and the performance increase with the number of OMP threads (Sect. \ref{s:mock_tests.scalability.ncores}). All the results given here correspond to the performance of \ASOHF on a Ryzen Threadripper 3960X processor with 24 physical cores.

\subsubsection{Scaling with $N_\mathrm{part}$ and $N_\mathrm{x}$}
\label{s:mock_tests.scalability.npart}
For assessing the scaling capabilities of the code with increasing number of particles and size of the base grid, we have replicated the set-up in Test 1 (Sect. \ref{s:mock_tests.field}) with varying number of particles.

\begin{figure*}
    \centering
    \includegraphics[width=\linewidth]{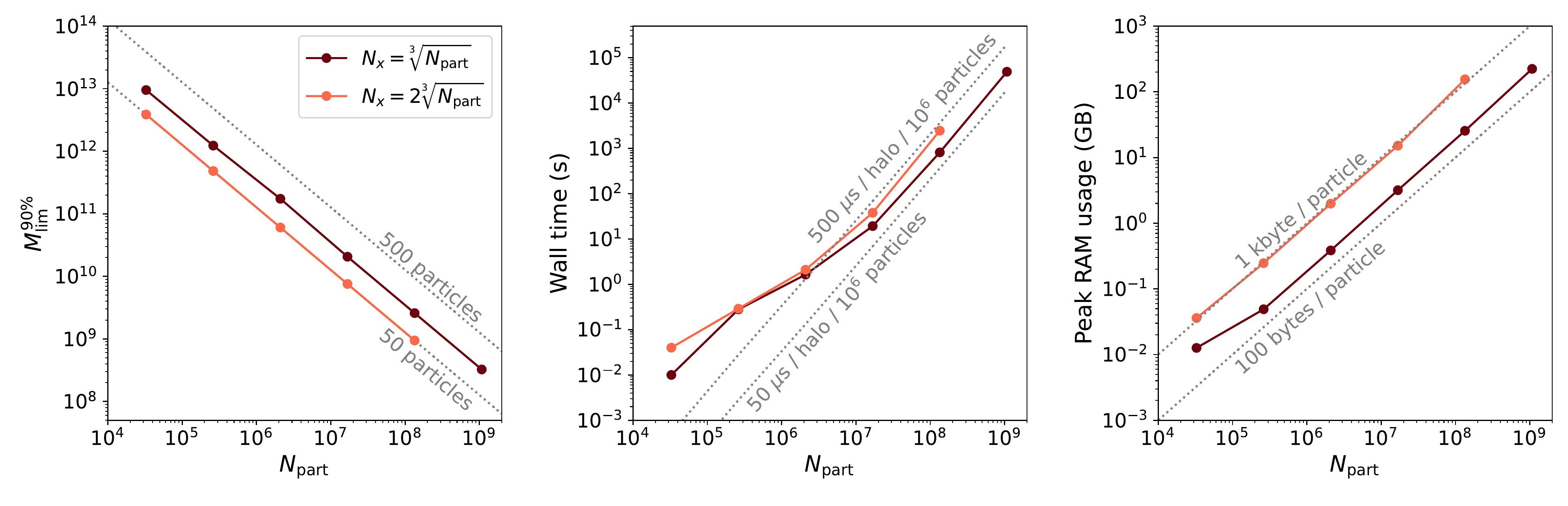}
    \caption{Summary of the results from the scalability test (Sect. \ref{s:mock_tests.scalability.npart}). \textit{Left panel:} Scaling of the $90\%$ completeness limit (in terms of mass) with number of particles in the domain. Gray, dashed lines correspond to a constant number of particles. Dark red (light red) lines represent the results for the two base grid sizes according to the legend. \textit{Middle panel:} Scaling of the wall time taken by \text{ASOHF} with number of particles. Gray, dashed lines correspond to a scaling $\propto N_\mathrm{part} N_\mathrm{haloes}^\mathrm{input}$. \textit{Right panel:} Scaling of the maximum RAM used \text{ASOHF} with number of particles. Gray, dashed lines correspond to a constant amount of RAM per particle.}
    \label{fig8}
\end{figure*}

To do so, we have first considered the case with the largest number of particles, $N_\mathrm{part}=1024^3$ (corresponding to a particle mass of $m_p^{1024^3} = 2.4 \times 10^6 M_\odot$, and have generated and randomly placed $366\,652$ haloes with masses over $M_\mathrm{min} = 50 m_p$. Then, we have realised this haloes catalogue with $N_\mathrm{part}=32^3$, $64^3$, ..., up to $1024^3$ DM particles. For each realisation, we only keep the haloes with mass greater than or equal to that corresponding to 50 particles. For each value of $N_\mathrm{part}$, we have tested $N_x = \sqrt[3]{N_\mathrm{part}}$ and $N_x = 2\sqrt[3]{N_\mathrm{part}}$,\footnote{Note that this is equivalent to setting the base grid cell size to the mean particle separation, and to half this value.} since it has been shown in \textit{Test 1} (Sect. \ref{s:mock_tests.field} that this latter value was able to recover all haloes. Regarding the AMR grid parameters, we have used $n_\mathrm{part}^\mathrm{refine}=8$, and the rest of parameters have been kept as in \textit{Test 1}. The detailed results are presented in Table \ref{tab:scalability}, and graphically summarised in Fig. \ref{fig8}.

\begin{table*}[]
    \centering
    \caption{Results from the scalability tests. Each row corresponds to a particular test, characterised by the number of particles ($N_\mathrm{part}$) and the base grid size ($N_x$). For each $N_\mathrm{part}$, the mass function is truncated at $M_\mathrm{min}^\mathrm{input}$ and $N_\mathrm{haloes}^\mathrm{input}$ are thus generated. For each run, we report the number of haloes detected by \ASOHF ($N_\mathrm{haloes}^\mathrm{ASOHF}$), the $90\%$ completeness limit (in mass and number of particles; $M_\mathrm{lim}^{90\%}$ and $n_\mathrm{part,lim}^{90\%}$, respectively), the execution (wall) time and the peak RAM usage. The last column gives the peak RAM per particle.}
    \begin{tabular}{cc||ccc|c|cccccc}
         \multicolumn{2}{c||}{Test} & $N_\mathrm{part}$ & $M_\mathrm{min}^\mathrm{input} \; (M_\odot)$ & $N_\mathrm{haloes}^\mathrm{input}$ &  $N_x$ & $N_\mathrm{haloes}^\mathrm{ASOHF}$ & $M_\mathrm{lim}^{90\%} \; (M_\odot)$ & $n_\mathrm{part,lim}^{90\%}$ & Wall time & Peak RAM & (bytes/part.)\\ \hline\hline
          \multirow{2}{*}{4.1} & a & \multirow{2}{*}{$32^3$} & \multirow{2}{*}{$3.86\times 10^{12}$} & \multirow{2}{*}{31} &   32 &    10 & $9.52\times 10^{12}$ & 123 & 10 ms &  12.6 MB &  385 \\
          & b & & &  &  64 & 31 & All & All &     40 ms &  36.0 MB           & 1098 \\
          \multirow{2}{*}{4.2} & a & \multirow{2}{*}{$64^3$} &  \multirow{2}{*}{$4.83\times 10^{11}$} &  \multirow{2}{*}{213} &   64 &  79 & $1.23\times 10^{12}$ & 127 & 280 ms &  49.0 MB          &  187 \\
          & b & & & &  128 &   214 & All & All & 290 ms & 247 MB           & 940  \\
          \multirow{2}{*}{4.3} & a & \multirow{2}{*}{$128^3$} & \multirow{2}{*}{$6.03\times 10^{10}$} & \multirow{2}{*}{1263} &  128 &   495 & $1.73\times 10^{11}$ & 143 & 1.64 s & 385 MB           &  184 \\
          & b & & & &  256 &  1263 & All & All & 2.1 s &      1.98 GB &  946 \\
          \multirow{2}{*}{4.4} & a & \multirow{2}{*}{$256^3$} & \multirow{2}{*}{$7.54\times 10^{9}$}& \multirow{2}{*}{8206} &  256 &  3099 & $2.06\times 10^{10}$ & 136 & 19.3 s & 3.18 GB &  190 \\
          & b &  & & &  512 &  8211 & All & All & 37.9 s &      15.1 GB &  903 \\
          \multirow{2}{*}{4.5} & a & \multirow{2}{*}{$512^3$} & \multirow{2}{*}{$9.43\times 10^{8}$} & \multirow{2}{*}{54\,476} &  512 & 20\,617 & $2.58\times 10^{9}$ & 137 & 13 min 38 s   &      25.5 GB &  190 \\
         & b &  & & & 1024 & 54\,495 & All & All & 40 min 39 s  &      155 GB & 1155  \\
         4.6 & a & $1024^3$ & $1.18\times 10^{8}$ & 366\,653 & 1024 & 136\,411 & $3.25 \times 10^{8}$ & 137 & 13 h 41 min & 223 GB & 208
    \end{tabular}
    \label{tab:scalability}
\end{table*}

The left panel in Fig. \ref{fig8} shows the scaling of the $90\%$ mass completeness limit with the number of particles, for the two base grid sizes ($N_x=\sqrt[3]{N_\mathrm{part}}$ in dark red, labelled `\textit{a}' in Table \ref{tab:scalability}; and $N_x=2\sqrt[3]{ N_\mathrm{part}}$ in light red, labelled `\textit{b}'; the same colour code is kept in the remaining panels). The gray, dashed lines are lines of constant number of particles (at fixed total mass in the box), so that it is explicitly shown that, using $N_x = \sqrt[3]{N_\mathrm{part}}$ (runs \textit{a}), the code is able to correctly identify barely all haloes comprising more than $\sim 140$ particles. For the runs \textit{b}, with $N_x = 2 \sqrt[3]{N_\mathrm{part}}$, all haloes have been detected and the completeness limit has been arbitrarily placed at the minimum mass of the input mass function (50 particles) for representation purposes. Nevertheless, it is worth stressing that, in real situations, where there is a LSS component surrounding haloes, it is not generally necessary to increase the base grid resolution beyond $\sqrt[3]{N_\mathrm{part}}$ to detect a larger amount of smaller haloes, although this may be useful for some particular applications (e.g., identifying haloes within voids).

The wall time taken by each of the runs is presented in the central panel of Fig. \ref{fig8}. The runs 4.1, 4.2 and 4.3, below a few million particles, last for less than a few seconds, and thus these results are biased with respect to the general scaling, since the measured times are likely to be dominated by input/output, system tasks and by the coarse time resolution of the profiling utility used. For large enough task sizes (i.e., for $N_\mathrm{part} \gtrsim 256^3$), we see that the wall time increases proportionally to the product $N_\mathrm{part} \times N_\mathrm{haloes}$, as shown by the dashed lines, which correspond to constant time per halo and particle. When increasing the base grid resolution from $N_x = \sqrt[3]{N_\mathrm{part}}$ to $N_x = 2 \sqrt[3]{N_\mathrm{part}}$, the scaling is kept, but the normalization of the relation increases in a factor of 2-3, mostly associated to the fact that more haloes are detected in this case (both real haloes, and spurious peaks over the interpolated density field, which get later discarded when using particles).

The right panel in Fig. \ref{fig8} depicts the peak memory requirements of the code. With the grid size fixed at $N_x =  \sqrt[3]{N_\mathrm{part}}$, \ASOHF requires around $\sim 180-200 \, \mathrm{bytes/particle}$, allowing to run the code on desktop-sized workstations for simulations with up to hundreds of millions of particles. When using twice the base grid resolution, these memory requirements increase up to $\sim 1\, \mathrm{kbyte/particle}$.

Last, we note that, as the number of particles increases, the scaling $\propto N_\mathrm{part} N_\mathrm{haloes}$ can greatly benefit from performing a domain decomposition for running \ASOHFns, as described in Sect. \ref{s:algorithm.additional.domdecomp}. This is especially useful in large domains, where the required overlaps amongst domains (which can be as low as the size of the largest halo expected) correspond to a negligible fraction of the volume. Thus, if decomposing the volume in $d$ domains, by dividing the number of particles and haloes in each domain, on average, by a factor of $d$ each, the CPU time reduces in a factor $d$. If all domains can be run concurrently, rather than sequentially, this amounts to an improvement of the wall time in a factor of $d^2$.

\subsubsection{Scaling with number of threads}
\label{s:mock_tests.scalability.ncores}
\begin{figure}
    \centering
    \includegraphics[width=0.85\linewidth]{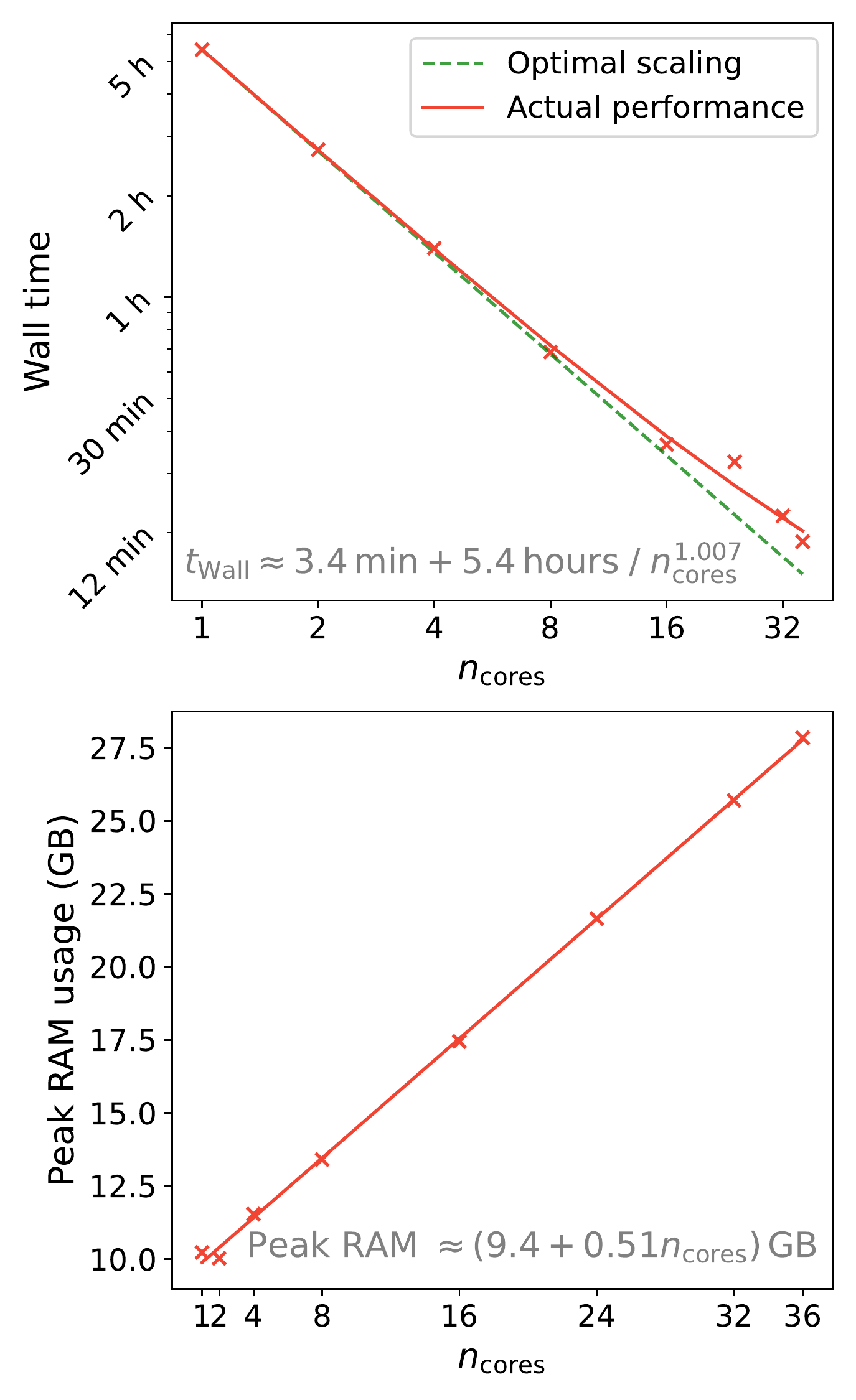}
    \caption{Parallel performance of \ASOHFns. \textit{Upper panel:} Scaling of the wall time of \ASOHF in Test 4.5a (see Table \ref{tab:scalability}), when varying the number of OMP threads. The red, solid line is a fit of the benchmark data (red crosses), while the green, dashed line corresponds to and optimal scaling, i.e., $t_\mathrm{Wall} \propto 1/n_\mathrm{cores}$. \textit{Lower panel:} Peak memory usage of \text{ASOHF} scaling with the number of OMP threads.}
    \label{fig9}
\end{figure}

To explore the performance gain, in terms of wall time and memory usage, of the OMP parallelisation scheme, we have repeated test 4.5a (see Table \ref{tab:scalability}) with $n_\mathrm{cores}=1$, $2$, $4$, $8$, $16$, $24$, and $32$ and $36$ OMP threads, using nodes equipped with two 18-core CPU Intel$^\text{\textregistered}$ Xeon$^\text{\textregistered}$ Gold 6154. The results are presented in Fig. \ref{fig9}.

The upper panel exemplifies the performance improvement when increasing the number of cores. Red crosses correspond to the actual performance of \ASOHF with varying number of threads. We have fitted these data to the functional form

\begin{equation}
    \Delta t_\mathrm{Wall} = \left( \Delta t_\mathrm{CPU} \right)_\mathrm{seq} + \frac{\left( \Delta t_\mathrm{CPU} \right)_\mathrm{par}}{n_\mathrm{cores}^\alpha},
\end{equation}

\noindent using a least-squares method, finding that the sequential part amounts for $\left( \Delta t_\mathrm{CPU} \right)_\mathrm{seq} \approx 3.4 \, \mathrm{min}$, while the parallel part would correspond to a CPU time of $\left( \Delta t_\mathrm{CPU} \right)_\mathrm{par} \approx 5.4 \, \mathrm{h}$. Being $\left( \Delta t_\mathrm{CPU} \right)_\mathrm{seq} \ll \left( \Delta t_\mathrm{CPU} \right)_\mathrm{par}$, the scaling of the code is close to optimal (null sequential part, which is represented by the green line in the upper panel of Fig. \ref{fig9}) for $n_\mathrm{cores} \lesssim 100$, which is a reasonably high number of threads available in typical shared memory nodes. The exponent $\alpha$ resulting from the fit is $\alpha =1.007 \pm 0.017 \approxeq 1$, which is consistent with the expected behaviour of the parallel part. It is interesting to note the absence of a significant performance gain when increasing from 16 to 24 cores, due to the fact that the system is comprised of two non-uniform memory access (NUMA) nodes, with 18 physical cores each. When increasing from 16 to 24 threads, the task can no longer be allocated in a single node, so that memory access outside the node is penalised. This is a well-known issue in shared-memory systems, and we advise the interested users to take the architecture of their system into account for optimal performance.

However, increasing the number of OMP threads implies an increased memory usage due to the necessity of replicating part of the data. The lower panel in Fig. \ref{fig9} displays the (linear) relation between the number of threads and the peak RAM used by the job. To end this section, we remind the reader that these results are dependent on the application and the configuration. For example, it is reasonable to expect a smaller memory penalty by increasing the number of threads in simulations of large volumes, where each halo contains only a small fraction of the particles in the domain, as opposed to these test cases where a halo contains nearly $25\%$ of the particles.


\section{Tests on real simulation data and comparison with other halo finders}
\label{s:simulation_tests}

Last, in order to evaluate the performance of \ASOHF on a cosmological simulation, we have tested our code on outputs from the public suite \texttt{CAMELS} \citep{Villaescusa-Navarro_2021, Villaescusa-Navarro_2022}. The \textit{Cosmology and Astrophysics with MachinE Learning Simulations} project consists of over 4000 simulations of $(25 h^{-1} \, \mathrm{Mpc})^3 \approx (37.25 \, \mathrm{Mpc})^3$ cubic domains, including DM-only and (magneto)hydrodynamic (MHD) simulations with star formation and feedback mechanisms, and covering a broad space of several cosmological and astrophysical parameters. Alongside with the simulation data, the public release also includes halo catalogues produced with three public halo finders, namely \texttt{SUBFIND} \citep{Springel_2001, Dolag_2009}, \texttt{AHF} \citep{Gill_2004, Knollmann_2009} and \texttt{ROCKSTAR} \citep{Rockstar_2013}, which we use here to examine how does \ASOHF compare to other well-known halo finders.

In particular, we make use of the simulation \texttt{LH-1} from the \texttt{IllustrisTNG} subset of \texttt{CAMELS}. These simulations were carried out using \texttt{Arepo} \citep{Springel_2010, Weinberger_2020}, a Tree+Particle-Mesh (TreePM) coupled to Voronoi moving-mesh MHD publicly available code for hydrodynamical cosmological simulations, including galaxy formation physics. The TreePM \citep{Bagla_2002} method implemented in \texttt{Arepo}, which combines a Particle-Mesh method for computing the large-range force with a more accurate tree code \citep{Barnes_1986} at short distances, allows these simulations to host a rich amount of small haloes and substructures even though the number of DM particles, $N_\mathrm{part}^\mathrm{DM}=256^3$, is modest. Indeed, the comoving gravitational softening length is as low as $2 \, \mathrm{kpc}$. While all \texttt{CAMELS} simulations correspond to flat Universes with baryon density parameter $\Omega_b=0.049$, Hubble constant $h \equiv H_0 / (100 \, \mathrm{km \, s^{-1} \, Mpc^{-1}})=0.6711$, and spectral index $n_s=0.9624$, the simulation we have chosen, name-coded \texttt{LH-1}, assumes matter density parameter $\Omega_m=0.3026$ and $\sigma_8=0.9394$ as the amplitude of the primordial fluctuations spectrum. These parameters imply a DM particle mass of $m_\mathrm{part} = 9.8 \times 10^7 \, M_\odot$.

We have run \texttt{ASOHF}, using only DM particles, on the most recent snapshot of this simulation (at redshift $z=0$), using a base grid with as many cells as particles ($N_x = 256$) and $n_\ell=6$ refinement levels, so that the peak resolution of the grid is $\Delta x_6 \approx 2.3 \, \mathrm{kpc}$, similar to the gravitational softening length of the simulation. The threshold particle number to mark a cell as refinable was set to 3, and the minimum patch size, to 14 cells. Density is interpolated with a kernel size determined by the local density, using 4 kernel levels. We discard all haloes resolved with less than 15 particles. In this configuration, \ASOHF detects $11\,794$ non-substructure haloes, the most massive of them with a DM mass of $7.44 \times 10^{13} M_\odot$, and 1263 substructures (including sub-substructures). The wall-time duration of the test has been below 4 minutes, using 8 cores in the same architecture described in Sect. \ref{s:mock_tests.scalability.npart}.

\begin{figure}
    \centering
    \includegraphics[width=\linewidth]{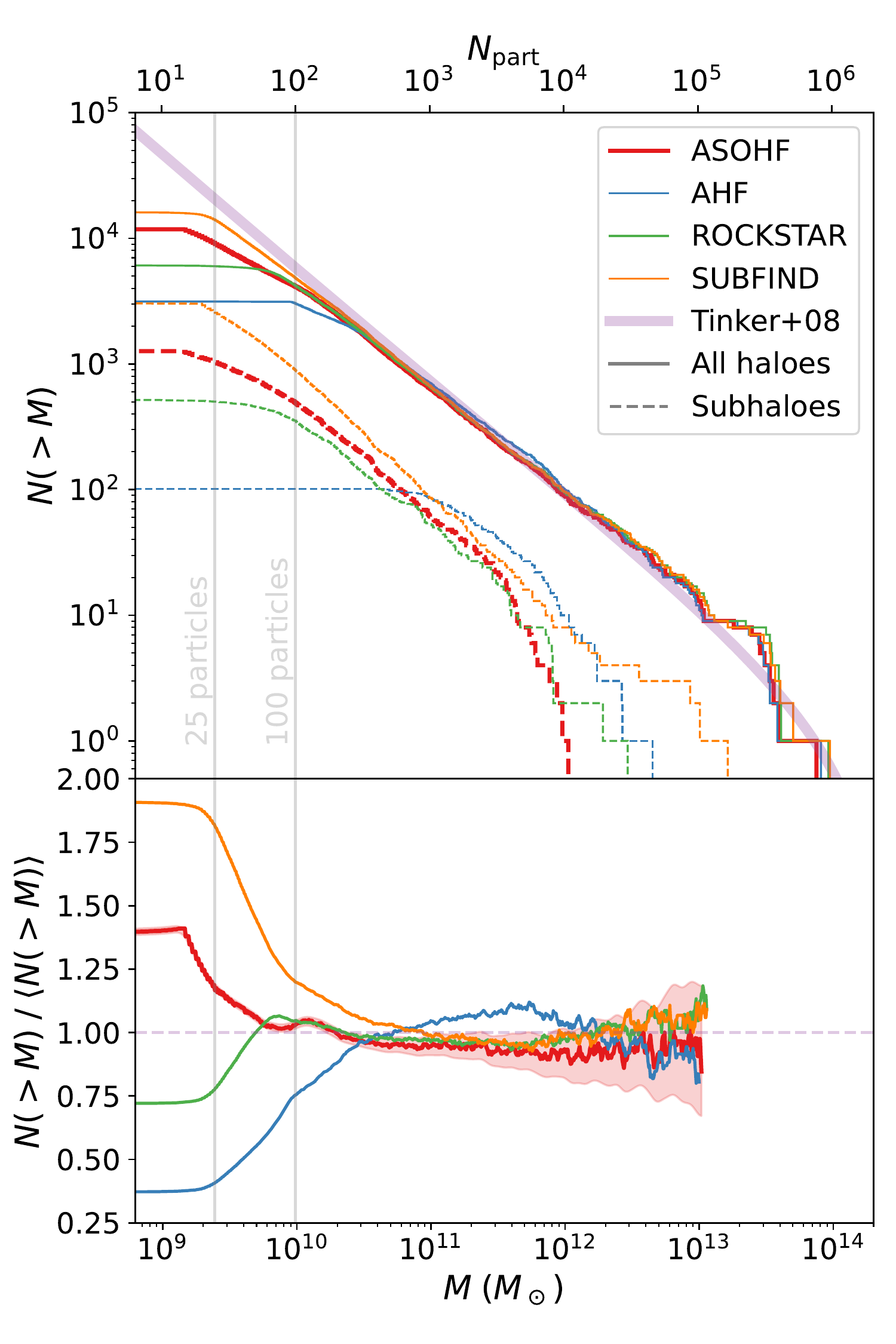}
    \caption{Comparison of the mass functions of the halo catalogues obtained by \ASOHF (red), \texttt{AHF} (blue), \texttt{Rockstar} (green) and \texttt{Subfind} (orange) in the \texttt{CAMELS} \texttt{Illustris-TNG} \texttt{LH-1} simulation at $z=0$. \textit{Upper panel:} Mass functions obtained by each code, in solid lines. The thick, purple line corresponds to a \cite{Tinker_2008} mass function at $z=0$ and with the cosmological parameters of the simulation, for reference. Dashed lines present, in the same colour scale, the mass function of subhaloes. \textit{Lower panel:} Mass function of non-substructure haloes, normalised by the geometric mean of this statistic for the four halo finders. The shadow on the red line (\ASOHFns) corresponds to $\sqrt{N}$ errors in halo counts, and are similar for the rest of finders.}
    \label{fig10}
\end{figure}

The halo mass function recovered by \texttt{ASOHF} is shown as the solid, thick red line in the upper panel of Fig. \ref{fig10}, together with the same quantity obtained from the catalogues of \texttt{AHF} (solid, blue), \texttt{Rockstar} (solid, green), and \texttt{Subfind} (solid, orange). For reference, the pale purple, thick line represents a reference, \cite{Tinker_2008} mass function with the cosmological parameters of the simulation. At first glance, all four halo finders yield comparable mass functions (both amongst them and with the reference one) for $N_\mathrm{part} \gtrsim 1000$, with some variations at the high-mass end due to the combination of small number counts and subtleties in the mass definitions. 

To allow a more precise comparison, the lower panel presents each of the mass functions, normalised to the geometric mean of all four finders, which we take as the baseline for comparison. In this plot, the red line (corresponding to \texttt{ASOHF}) displays the Poisson ($\sqrt{N}$) confidence intervals as the red shadowed area. The magnitude of these uncertainties is similar for the rest of lines. \ASOHFns, \texttt{Subfind} and \texttt{Rockstar} match reasonably well inside their respective confidence intervals, while \texttt{AHF} shows a small, $\gtrsim 15 \%$ excess with respect to them at $N_\mathrm{part} \sim 10^4$. Interestingly, \texttt{Rockstar} and \ASOHF present the largest similarities in their mass function, matching each other within a few percents all the way down to $50-100$ particles, when \ASOHF starts to identify a larger amount of structures. Compared to them, \texttt{Subfind} starts to present a larger abundance of haloes below $N_\mathrm{part} \lesssim 1000$, mostly driven by the larger amount of substructures; while \texttt{AHF} halo counts stall below a few hundred particles. 

Dashed lines in the upper panel of Fig. \ref{fig10} present the substructure mass functions for the four halo finders, using the same colour code. In this case, substructure mass functions are not so easily comparable and thus differences amongst finders are exacerbated \citep[see, e.g.,][]{Onions_2013}. In particular, \texttt{Subfind} finds the largest amount of substructure, nearly 2.5 times as many as \ASOHFns, 6 times more than \texttt{Rockstar} and 30 times above \texttt{AHF}. Even in the high-mass end of the subhalo mass function there are important differences, with \texttt{Subfind} having the largest and \ASOHF the smallest mass of their respective most massive substructure. This is mainly due to the fundamentally different definitions of substructure. \ASOHF choice of using the Jacobi radius, which approximately delimits the region where the gravitational attraction of the substructure is stronger than that of the host, is more restrictive than other definitions, like using density contours traced by the AMR grid (\texttt{AHF}), reducing the 6D linking-length (\texttt{Rockstar}) or finding the saddle point of the density field (\texttt{Subfind}). This more stringent definition of the substructure extent may, indeed, also be behind the fact that \ASOHF recovers less substructure than \texttt{Subfind}, since many of the small substructures identified by the latter may contain less than 15 particles using our more stringent definition based on the Jacobi radius.

\begin{figure*}
    \centering
    \includegraphics[width=\linewidth]{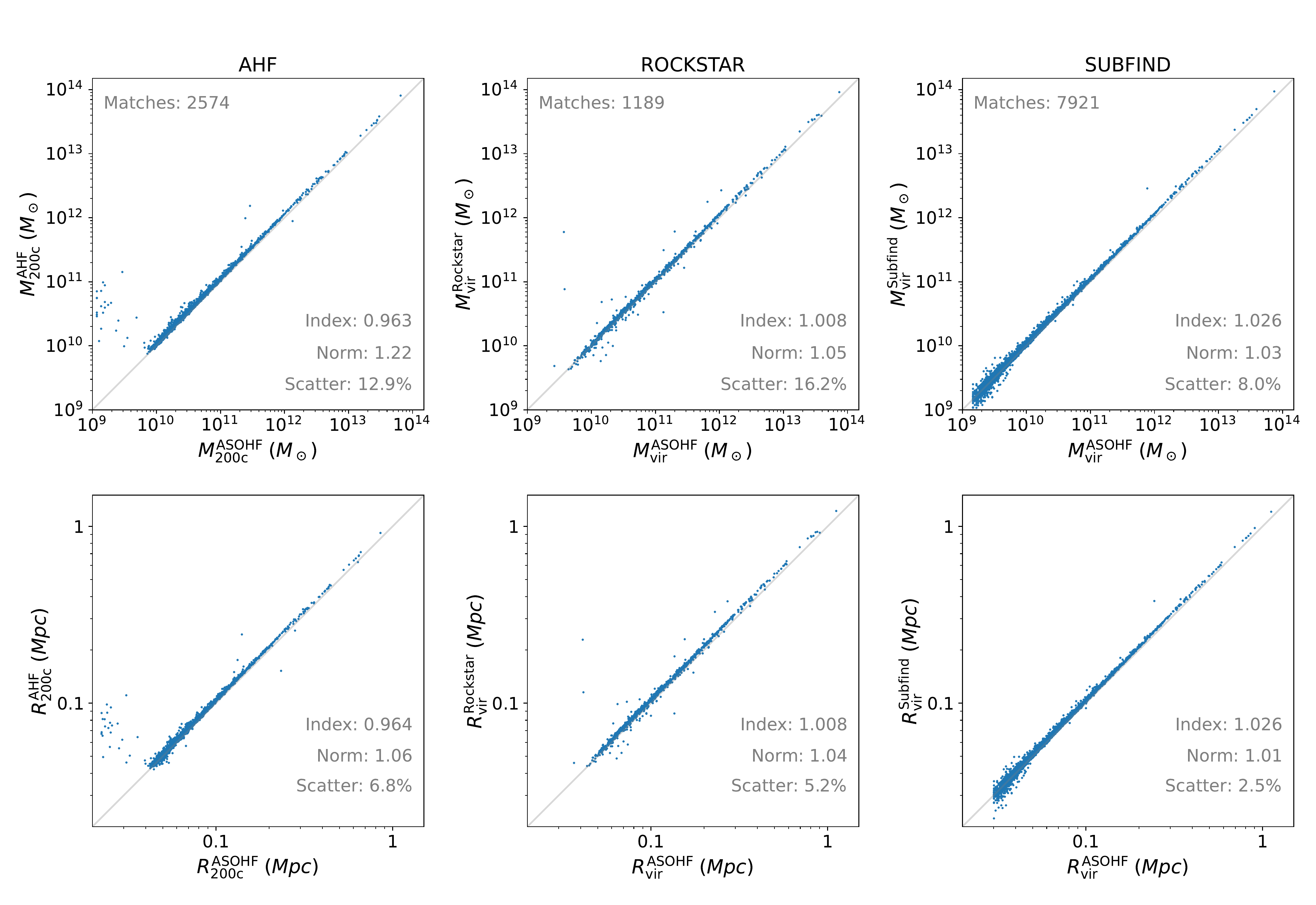}
    \caption{Comparison of the main properties (masses, \textit{top row}; and radii, \textit{bottom row}) of the matched haloes between \ASOHF and \texttt{AHF} (left column), \texttt{Rockstar} (middle column) and \texttt{Subfind} (right column). For the two latter we compare virial radii and masses, while for the former we use $R_\mathrm{200c}$ and $M_\mathrm{200c}$, since the \texttt{AHF} catalogues given in the public release of \texttt{CAMELS} have used this overdensity for marking the boundary of the halo.}
    \label{fig11}
\end{figure*}

\begin{table*}[]
    \centering
    \small
    \caption{Summary of the main features of the comparison between \ASOHFns, \texttt{AHF}, \texttt{Rockstar} and \texttt{Subfind}. Left to right, columns contain the total number of haloes detected ($N_\mathrm{haloes}$), total number of substructures ($N_\mathrm{subs}$), mass of the largest halo ($M_\mathrm{max}$), maximum number of substructures of a single halo ($\max (n_\mathrm{subs})$), and normalisation, index and scatter ($N_X$, $\alpha_X$, and $s_X$; as defined in Eqs. \ref{eq:powerlaw_comparison} and \ref{eq:scatter_comparison}) of the radii and mass comparison between each finder and \ASOHFns.}
    \begin{tabular}{c|cccc|ccc|ccc}
        \multirow{2}{*}{Halo finder} & \multicolumn{4}{c|}{Halo statistics} & \multicolumn{3}{c|}{Radii fit} &  \multicolumn{3}{c}{Masses fit}\\
         & $N_\mathrm{haloes}$ & $N_\mathrm{subs}$ & $M_\mathrm{max} \; (10^{13} M_\odot)$ & $\max (n_\mathrm{subs})$ & $N_R$ & $\alpha_R$ & $s_R$ & $N_M$ & $\alpha_M$ & $s_M$ \\ \hline
        \ASOHF & 11794 & 1263 & 7.44 & 122 &- & -& -& -& -& -\\
        \texttt{AHF} & 3141 & 101 & 8.07 & 7 & $1.0595(33)$ & $0.9639(45)$ & $6.8\%$& $1.219(13)$ & $0.9627(48)$ & $12.9\%$\\
        \texttt{Rockstar} & 6089 & 519 & 9.14 & 50 & $1.0369(43)$ & $1.0077(38)$ & $5.2\%$& $1.051(12)$ &$1.0077(38)$ & $16.2\%$\\
        \texttt{Subfind} & 16129 & 3037 & 9.38 & 343 & $1.00968(33)$ & $1.02632(61)$ & $2.5\%$ & $1.03081(96)$ &$1.02632(61)$ & $8.0\%$\\
    \end{tabular}
    \label{tab:comparison}
\end{table*}

Focusing on the main properties of haloes, in Fig. \ref{fig11} we present the comparison of \ASOHF radii and masses to the ones obtained by \texttt{AHF} (left column), \texttt{Rockstar} (middle column) and \texttt{Subfind} (right column). For each halo finder, we have first matched its catalogue to the one produced by \ASOHFns. Here, we only focus on haloes and exclude substructure because, as mentioned above, substructure definitions of each algorithm are fundamentally different. The results are also summarised in Table \ref{tab:comparison}. 

Surprisingly, and despite their agreement regarding the mass functions, we are only able to match $\sim 20\%$ of \texttt{Rockstar} haloes to \ASOHFns's,\footnote{This is due to some systematic off-centrering between the halo centres reported by \texttt{Rockstar}, and those reported by the rest of finders (as well as \ASOHFns) in the \texttt{Camels} suite, whose origin falls beyond the scope of this paper.} in contrast to $\sim 80\%$ of matches when comparing to \texttt{AHF} and \texttt{Subfind}. For each property, $X$, to compare, we have fitted the matched data to a power law of the form,

\begin{equation}
    \log \frac{X^\mathrm{finder}}{X^*} = \log N + \alpha \log \frac{X^\mathrm{ASOHF}}{X^*} 
    \label{eq:powerlaw_comparison}
\end{equation}

\noindent where $N$ is the normalisation and $\alpha$ is the index (both unity if the finders yielded identical results). $X^*$ is a normalisation for the quantities, which we pick as the median over the \ASOHF sample: $R^* = 43.2 \, \mathrm{kpc}$ and $M^* = 5.18 \times 10^9 \, M_\odot$. For each fit, we compute the scatter as the RMS of the relative residuals between the data points and the fit,

\begin{equation}
    s = \sqrt{\frac{1}{N_\mathrm{matched}} \sum_{i=1}^{N_\mathrm{matched}} \left(1-\frac{\hat X^\mathrm{finder}(X^\mathrm{ASOHF})}{X^\mathrm{finder}}\right)^2},
    \label{eq:scatter_comparison}
\end{equation}

\noindent where $\hat X^\mathrm{finder}(X^\mathrm{ASOHF})$ is the fitting function from Eq. \ref{eq:powerlaw_comparison} with the best parameters obtained by least-squares estimation.

The largest discrepancies with \ASOHF are seen when comparing it to \texttt{AHF}, which overall finds $\sim 6\%$ larger radii when compared to \ASOHFns. This may be well due to the fact that \texttt{AHF} uses all particles (including gas and stars) to determine the spherical overdensity radius, while we only use DM particles. This normalisation offset in radii propagates to a $\sim 22\%$ offset in masses. There is a significant amount of scatter in radii and masses at the low-mass end, with a population of matched haloes with \texttt{AHF} radii a few times that of \ASOHFns, which is not seen when comparing with the rest of halo finders.

Although only $\sim$1200 haloes are matched between \texttt{Rockstar} and \ASOHF catalogues, their virial radii and masses match tightly, with $\sim 4\%$ and $5\%$ offsets, respectively. A few outliers increase the overall scatter figure, which reaches $\sim 5\%$ and $\sim 16\%$ for radii and masses, respectively, but excluding them, the relation is quite tight.

Strikingly, despite their very different natures, \texttt{Subfind} results are the most compatible with \ASOHF among the halo finders considered. With almost 8000 matched haloes, the normalisations, as defined in Eq. \ref{eq:powerlaw_comparison}, only show a $1\%$ offset in radii, and $3\%$ offset in mass; while the scatter, dominated by the low-mass haloes, is of $2.5\%$ and $8\%$, respectively.

Overall, and in spite of the difficulties in comparing results of different halo finders in a halo-to-halo basis, which has been thoroughly explored in the literature (c.f. \citealp{Knebe_2011, Onions_2012, Knebe_2013}), these analyses show that \ASOHF is capable of providing results in robust general agreement with other widely-used halo finders.

\subsection{Results including stellar particles}
\label{s:simulation_tests.stars}

By the last snapshot, at $z=0$, the simulation contains $227\,454$ stellar and black hole (BH) particles, accounting for $\sim 2 \permil$ of the DM mass in the computational domain. While the mass in stars is too little to make a significant impact on the process of halo finding (by enhancing DM density peaks that would otherwise be too small to be detected), we can still demonstrate the ability of \ASOHF to characterise stellar haloes.

We have run \ASOHF on the DM+stellar particles of the same simulation and snapshot considered above, keeping the same values for the mesh creation and DM halo finding parameters. Regarding the stellar finding parameters, which as discussed in Sect. \ref{s:algorithm.stellar} do not have a severe impact on the resulting galaxy catalogues, we have kept all stellar haloes with more than $N^\mathrm{stellar}_\mathrm{min}=15$ stellar particles inside the half-mass radius, and we have preliminary cut the stellar haloes whenever density increased by more than a factor $f_\mathrm{min}=5$ from the profile minimum, there was a radial gap of more than $\ell_\mathrm{gap}=10 \, \mathrm{kpc}$ without any stellar particle or stellar density fell below the background density, $\rho_B(z)$ ($f_B=1$).

\begin{figure}
    \centering
    \includegraphics[width=\linewidth]{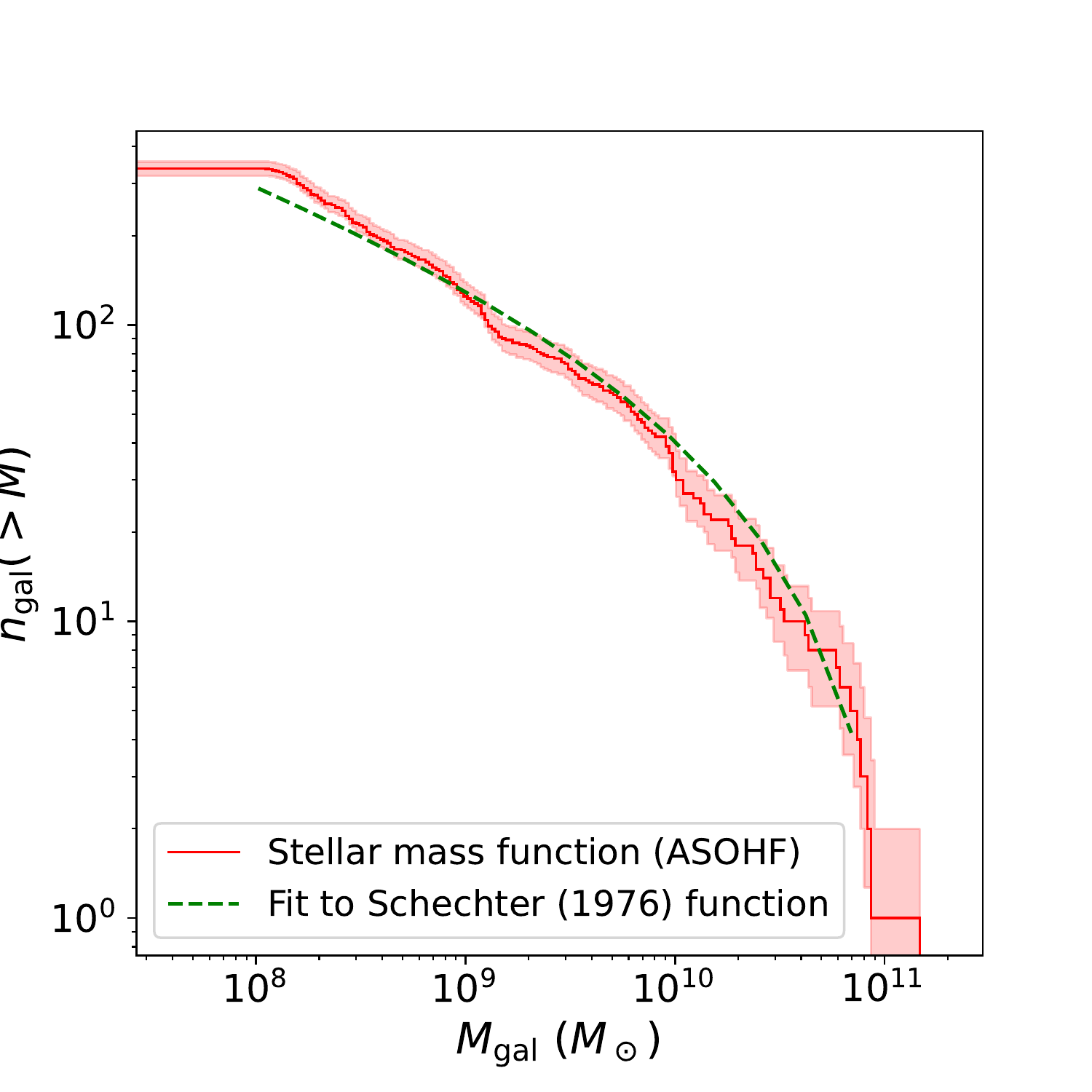}
    \caption{Mass function of stellar haloes found in the simulation from \texttt{CAMELS} at $z=0$ (red line). The mass of the galaxy in the horizontal axis, $M_\mathrm{gal}$, refers to the stellar mass inside the half-mass radius of the stellar halo. The red shading represents the Poissonian, $\sqrt{N}$ uncertainties. The green, dashed lines corresponds to the best least-squares fit of the differential mass function to a \cite{Schechter_1976} parametrisation.}
    \label{fig12}
\end{figure}

In this configuration, \ASOHF identifies 337 stellar haloes, with masses ranging between $1.1\times 10^8 M_\odot$ and $1.5 \times 10^{11} M_\odot$, and whose cumulative mass distribution is shown in Fig. \ref{fig12} by a red line (here, $M_\mathrm{gal}$ makes reference to the stellar mass inside the stellar half-mass radius, defined in Sect. \ref{s:algorithm.stellar}). The green, dashed line presents a fit to a \cite{Schechter_1976} function. The fit was performed over the differential mass-function, and obtained a chi-squared per degree of freedom $\chi^2_\nu \approx 1.8$ (computed assuming Poissonian errors for number counts), implying that the fit is not inconsistent with a \cite{Schechter_1976} mass function.

\begin{figure}
    \centering
    \includegraphics[width=1.1\linewidth]{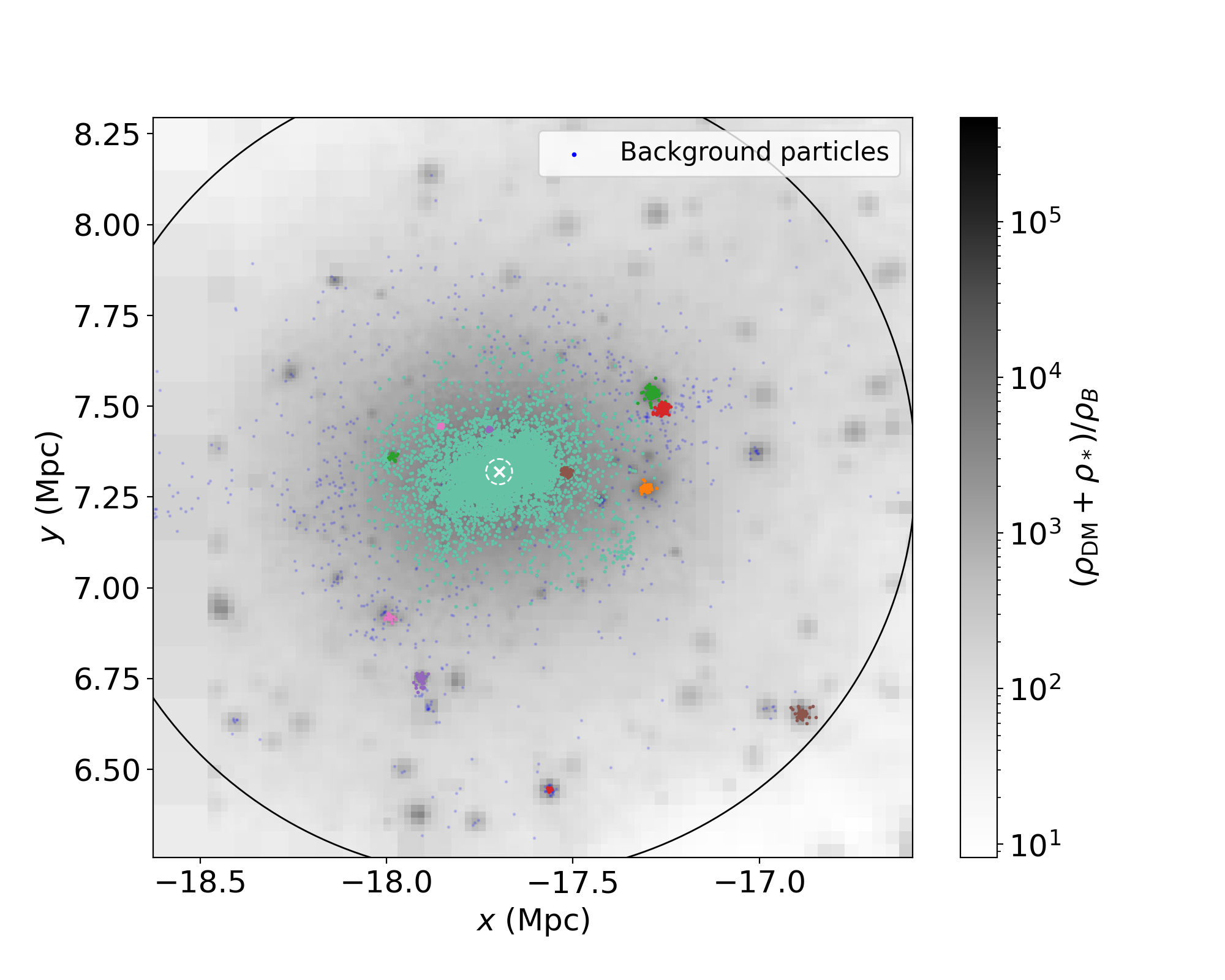}
    \caption{Example of the stellar halo finding capabilities of \ASOHFns. The background, greyscale colourmap is a projection of DM+stellar density around the most massive halo in the simulation, with the solid, black line marking its virial radius. The coloured dots correspond to stellar particles, with each colour corresponding to a different stellar halo. The dark blue dots are the particles inside the virial volume of the DM halo which do not belong to any identified stellar structure. For the central galaxy, the white cross and circle represent, respectively, the stellar density peak (which we regard as centre) and the half-mass radius of the stellar halo.}
    \label{fig13}
\end{figure}

An example of such identification is represented graphically in Fig. \ref{fig13}. In this figure, the background colour encodes, in grey scale, the DM+stellar projected density, as computed by \ASOHFns, around the most massive DM halo in the simulation (which is a $\sim 8 \times 10^{13} M_\odot$ group). Each colour represents all the particles of a given stellar halo. A central halo, represented in turquoise, dominates the stellar mass budget. This halo has a half-mass stellar radius of $\sim 36 \, \mathrm{kpc}$, whose extent is represented as the dashed, white circle in the figure, around a white cross marking the stellar density peak. Around it, \ASOHF detects ten satellite haloes, each represented in a different colour. In dark blue, we have represented all the particles inside the virial volume of the main DM halo which do not belong to any stellar halo. Although a few groups of stars on top of some DM haloes, not identified as galaxies, are seen, it can be checked that these objects correspond to poor stellar haloes, with less than 15 stellar particles, and are thus discarded by \ASOHFns.

\begin{figure*}
    \centering
    \includegraphics[width=\linewidth]{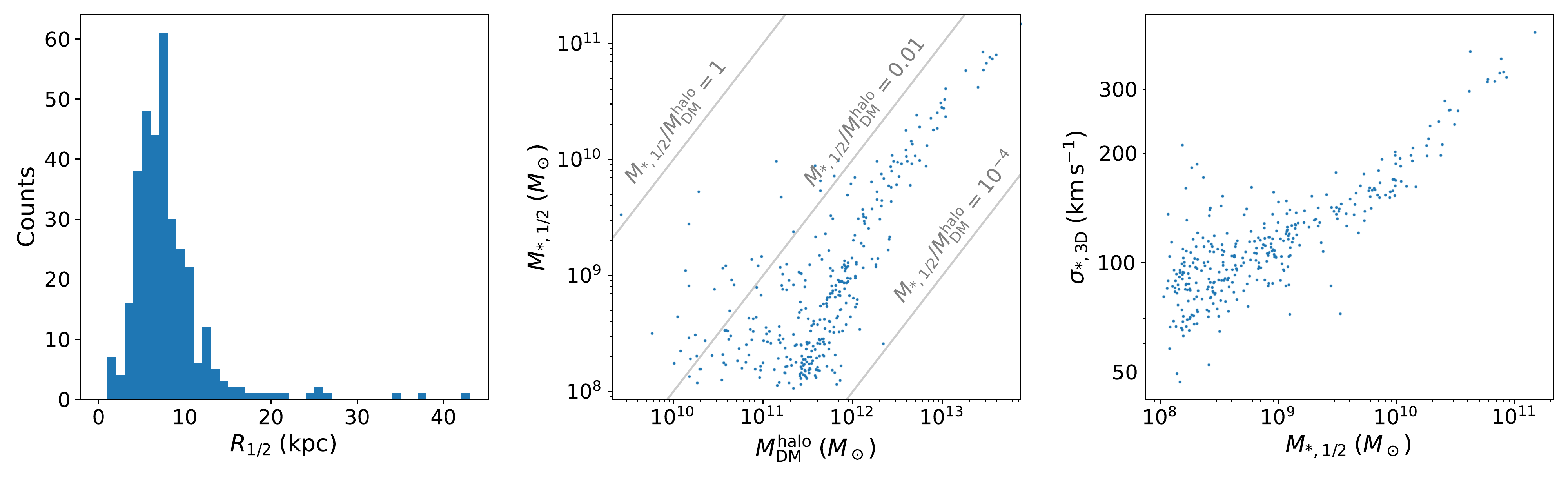}
    \caption{Summary of some of the properties of the stellar haloes identified by \ASOHF in the \texttt{CAMELS}, \texttt{IllustrisTNG LH-1} simulation at $z=0$. \textit{Left-hand panel:} Distribution of half-mass radii. \textit{Middle panel:} stellar mass to DM halo mass relation. Diagonal lines correspond to constant stellar fractions. \textit{Right-hand panel:} 3D velocity dispersion to stellar mass relation.}
    \label{fig14}
\end{figure*}

We summarise some of the properties of the stellar haloes catalogue in Fig. \ref{fig14}. The left-hand panel contains a histogram of the half-mass stellar radii. The distribution of galaxy sizes peaks at around $(6-8)\mathrm{\, kpc}$, while a handful of haloes as large as $R_{1/2} \gtrsim 20 \, \mathrm{kpc}$ are found, most typically associated to massive DM haloes. Although more rare, some of them correspond to small ($\sim 10^{11} M_\odot$) DM haloes with an extended stellar component.

In the middle panel of Fig. \ref{fig14}, we present the relation of stellar mass inside the half-mass radius (vertical axis) to DM host halo mass (horizontal axis). For reference, we indicate some constant stellar fractions in gray lines. At low stellar mass ($M_{*,1/2}\lesssim 2 \times 10^9$), we find a large scatter between stellar and DM halo masses; while the relation becomes more tight at high stellar halo masses, approaching but not reaching $1\%$. The particularities about these results may have to do with the particular \texttt{IllustrisTNG} feedback scheme implemented in the simulation and numerical resolution, and are out of the scope of this work.

Last, the right-hand panel of Fig. \ref{fig14} contains the stellar, 3D-velocity dispersion correlated to the stellar mass of the halo. There is a clear increasing trend in this relation, as more massive haloes (if virialised) have larger kinetic energy budgets, or are dynamically hotter. This result provides a dynamical confirmation of the fact that our galaxy identification scheme is targetting \textit{bona fide} stellar haloes, and not spurious groups of particles. 

In sum, with these examples we show the capabilities of \ASOHFns, not only as DM halo finder but also as a galaxy finder for cosmological simulations.

\section{Conclusions}
\label{s:conclusions}

In this paper, we have introduced and discussed a revamped version of the dark matter halo finder \ASOHFns, which constitutes a profound revision of most of the features of the original version, presented over a decade ago by \cite{Planelles_2010}. The ever-increasing trend in cosmological simulation sizes and resolution and, therefore, abundance and richness of structures at a large range of scales, demanded to revisit several aspects related to the halo finding process. Being the basic building blocks of the LSS of the Universe, an exceptionally precise characterisation of DM haloes must be a \textit{sine qua non} condition for bridging the gap between simulations and observations and bringing cosmological predictions from simulations below the percent level \cite[e.g.][for several reviews about cluster cosmology]{Borgani_2008, Clerc_2022}, especially when the smallest scales are involved \citep[e.g.][and references therein]{Vogelsberger_2020}. Related to this, as the baryonic processes leading to galaxy formation are being profusely analysed in numerical works \citep[e.g.][for some large galaxy formation simulation projects]{Murante_2010, Iannuzzi_2012, Vogelsberger_2014, Schaye_2015, Kaviraj_2017, Pillepich_2018, Villaescusa-Navarro_2021}, it is of utmost importance to devise numerical techniques to accurately describe them.

Amongst the main improvements to the halo finding procedure, which have been assessed in a battery of complex, idealised tests and in a cosmological simulation, we can summarise the following points, which can be useful to generalise, not only to \ASOHFns, but to halo finders in general (especially, to those based on configuration-space).

\begin{enumerate}
    \item The density interpolation procedure is, naturally, critical to the procedure of halo finding, since an overly smoothed interpolation would accidentally smear true density peaks, thus losing small-scale structures, while a too sharp density interpolation (e.g., applying a CIC or TSC at the resolution of each grid) increases dramatically the computational cost of the process by producing a large amount of spurious peaks due to sampling noise. We find that a compromise between these extremes is achieved by spreading particles in a cloud, whose size can be tuned either by local density or by particle mass (i.e., by the volume sampled by the particle in the initial conditions), but is irrespective of the grid level. Nevertheless, since any interpolation will tend to be more diffuse than the particle distribution itself, it is generally harmless to also consider density peaks above a certain fraction of $\Delta_\mathrm{vir}(z)$ as candidates for halo centres.
    
    Due to the finite resolution of the grid, in which one cell can contain thousands of particles even at the finest levels for very cuspy haloes, the implementation of a recentring scheme (using information at finer levels, and ultimately using the particle distribution to iteratively relocate the local maximum of the density field) is crucial for avoiding miscentring. Preventing this miscentring is critical for describing the properties of the inner regions of haloes.
    
    \item While the local escape velocity unbinding, implemented in many halo finders, is successful in removing physically unbound particles, the fact that a conservative threshold (e.g., $2 v_\mathrm{esc}$) is normally used can make it insufficient in some occasions. This can be improved by an additional unbinding scheme in pure velocity-space, which proves useful to disentangle dynamically-unrelated particles from the halo (see the examples in Sect. \ref{s:mock_tests.unbinding}, especially Fig. \ref{fig7}).

    \item The new version of \ASOHF implements a totally new scheme for looking for substructure, which is emancipated from the search of isolated haloes. In our experiments, we find that delimiting substructures by their density profile (e.g., placing a radial cut when the spherically-averaged density profile increases) leads to too generous radii, which therefore contaminate the mean velocity of the halo and can degrade the performance of unbinding and even miss the halo. In \ASOHFns, density peaks located within the volume of previously identified haloes, at finer levels of refinement, are characterised by their Jacobi radius, which approximately delimits the region where the gravitational force towards the substructure centre is dominant with respect to that of the host. Besides its clear physical motivation, this (typically more stringent) radius together with the improved unbinding scheme helps to avoid the bias in the properties of the halo.

    \item Using the DM haloes, \ASOHF is able to identify and characterise stellar haloes, i.e. galaxies. While the starting point for such identification is the catalogue of DM haloes and subhaloes, the stellar haloes are subsequently treated as independent objects. In particular, they can have a different centre, radial extent, dynamics, etc. than the DM halo they are built onto.

    \item We have implemented a domain decomposition for \ASOHF that enables the analysis of large simulations, which would otherwise not fit in memory. Incidentally, as discussed in Sects. \ref{s:algorithm.additional.domdecomp} and \ref{s:mock_tests.scalability.npart}, the domain decomposition also reduces importantly the CPU time. Furthermore, the procedure is implemented by external, automated scripts, and each domain can be treated as a separate job (requiring no communication). This means different domains can be either run concurrently (so that the wall time is reduced by a factor equal to the number of cores) or sequentially. The resulting outputs for each domain can be merged by an independent script, so that the whole process is transparent to the user.

    \item By saving the particle IDs of all the bound particles of each halo, the merger tree can be easily built once the results have been got for (at least) a pair of code outputs. This process is dramatically accelerated by pre-sorting the lists of particles by ID, and thus intersecting them in $\mathcal{O}(N)$ operations, instead of $\mathcal{O}(N^2)$. Building the merger tree as a separate task from the halo finding procedure has several benefits, such as avoiding the need to overload the memory with all the information of previous iterations or the ability to connect a halo with a progenitor skipping iterations, if it has not been found in the immediately previous iteration. Naturally, there are also drawbacks of this approach: for instance, the inability to incorporate the merger tree information into the halo finding process itself, such as other modern finders do (e.g., \texttt{HBT+}, \citealp{Han_2018}).
\end{enumerate}

Through a battery of tests, described in Sect. \ref{s:mock_tests}, we have shown \ASOHF capabilities of recovering virtually all haloes and subhaloes in a broad range of masses in idealised, yet complex, set-ups. When compared to other publicly available halo finders, such as \texttt{AHF}, \texttt{Rockstar} or \texttt{Subfind}, \ASOHF produces comparable results, in terms of halo mass functions and properties of haloes; providing excellent results in terms of substructure identification (in the sense that it is able to identify a large amount of substructures, while using a tight, physically-motivated definition of substructure extent), which is commonly a tough task for SO halo finders, when compared to 3D-FoF, but especially to 6D-FoF finders.

Regarding computational performance, the new version of \ASOHF has been rewritten to be extremely efficient, both in terms of speed, parallel scaling and memory usage. As discussed in Sect. \ref{s:mock_tests.scalability.ncores}, the parallel behaviour of the code scales nearly ideally, which allows a large performance gain using shared-memory platforms with large number of cores. Moreover, the optimization of the code allows to analise individual snapshots from simulations of $\gtrsim 10^7$ particles within tens of seconds and using a few GB of RAM; and simulations with $\gtrsim 10^8$ particles within tens of minutes and using below $30 \, \mathrm{GB}$ of RAM in any case. 

When large domains with even larger number of particles are involved, the time taken by \ASOHF can be further improved by the domain decomposition strategy, which involves no communication between different domains until a last step in which catalogues are merged. As an example, for a simulation with $\sim 10^{10}$ particles within a cubic domain of sidelength $500 \, \mathrm{Mpc/h}$, where $\sim 15 \times 10^6$ haloes are expected to form by $z=0$, scaling the benchmarks characterised in \ref{s:mock_tests.scalability.npart} we estimate that, performing a decomposition in $d = 4^3$ sub-domains, \ASOHF could run over all the domains (sequentially, i.e., one domain at a time) in $\sim 93 \,\mathrm{h}$ of wall time, in a 24-core processor like the one used for the Tests in Sect. \ref{s:mock_tests.scalability.npart}. If 8 of such nodes are concurrently available in a distributed memory architecture, this would take $\lesssim 12 \, \mathrm{h}$.

The implementation of \ASOHFns, together with the several utilities and scripts mentioned in this paper, is publicly available in and documented\footnote{See links to the code repository and documentation at the beginning of Sec. \ref{s:algorithm}.}.

\begin{acknowledgements}
We thank the anonymous referee for his/her constructive comments, which have improved the presentation of this work. This work has been supported by the Spanish Ministerio de Ciencia e Innovación (MICINN, grant PID2019-107427GB-C33) and by the Generalitat Valenciana (grant PROMETEO/2019/071). D.V. acknowledges support from Universitat de València through an \textit{Atracció de Talent} fellowship. Part of the tests have been carried out with the supercomputer Lluís Vives at the Servei d’Informàtica of the Universitat de València. This research has made use of the following open-source packages: \texttt{NumPy} \citep{Numpy}, \texttt{SciPy} \citep{Scipy}, \texttt{Matplotlib} \citep{Matplotlib}, and \texttt{Colossus} \citep{Diemer_2018}.
\end{acknowledgements}

\bibliographystyle{aa} 
\bibliography{asohf}

\begin{thebibliography}{60}
\expandafter\ifx\csname natexlab\endcsname\relax\def\natexlab#1{#1}\fi

\bibitem[{{Angulo} \& {Hahn}(2022)}]{Angulo_2022}
{Angulo}, R.~E. \& {Hahn}, O. 2022, Living Reviews in Computational
  Astrophysics, 8, 1

\bibitem[{{Bagla}(2002)}]{Bagla_2002}
{Bagla}, J.~S. 2002, Journal of Astrophysics and Astronomy, 23, 185

\bibitem[{{Barnes} \& {Hut}(1986)}]{Barnes_1986}
{Barnes}, J. \& {Hut}, P. 1986, \nat, 324, 446

\bibitem[{{Behroozi} {et~al.}(2013){Behroozi}, {Wechsler}, \&
  {Wu}}]{Rockstar_2013}
{Behroozi}, P.~S., {Wechsler}, R.~H., \& {Wu}, H.-Y. 2013, \apj, 762, 109

\bibitem[{{Binney} \& {Tremaine}(1987)}]{Binney_Tremaine}
{Binney}, J. \& {Tremaine}, S. 1987, {Galactic dynamics}, {Princeton Series in
  Astrophysics} ({Princeton University Press})

\bibitem[{{Borgani}(2008)}]{Borgani_2008}
{Borgani}, S. 2008, in A Pan-Chromatic View of Clusters of Galaxies and the
  Large-Scale Structure, ed. M.~{Plionis}, O.~{L{\'o}pez-Cruz}, \& D.~{Hughes},
  Vol. 740 (Springer), 24

\bibitem[{{Bryan} \& {Norman}(1998)}]{Bryan_Norman_98}
{Bryan}, G.~L. \& {Norman}, M.~L. 1998, \apj, 495, 80

\bibitem[{{Ca{\~n}as} {et~al.}(2019){Ca{\~n}as}, {Elahi}, {Welker}, {del P
  Lagos}, {Power}, {Dubois}, \& {Pichon}}]{Canas_2019}
{Ca{\~n}as}, R., {Elahi}, P.~J., {Welker}, C., {et~al.} 2019, \mnras, 482, 2039

\bibitem[{{Clerc} \& {Finoguenov}(2022)}]{Clerc_2022}
{Clerc}, N. \& {Finoguenov}, A. 2022, arXiv e-prints, arXiv:2203.11906

\bibitem[{{Cole} \& {Lacey}(1996)}]{Cole_1996}
{Cole}, S. \& {Lacey}, C. 1996, \mnras, 281, 716

\bibitem[{{Davis} {et~al.}(1985){Davis}, {Efstathiou}, {Frenk}, \&
  {White}}]{Davis_1985}
{Davis}, M., {Efstathiou}, G., {Frenk}, C.~S., \& {White}, S.~D.~M. 1985, \apj,
  292, 371

\bibitem[{{Diemand} {et~al.}(2006){Diemand}, {Kuhlen}, \&
  {Madau}}]{Diemand_2006}
{Diemand}, J., {Kuhlen}, M., \& {Madau}, P. 2006, \apj, 649, 1

\bibitem[{{Diemer}(2018)}]{Diemer_2018}
{Diemer}, B. 2018, \apjs, 239, 35

\bibitem[{{Dolag} {et~al.}(2009){Dolag}, {Borgani}, {Murante}, \&
  {Springel}}]{Dolag_2009}
{Dolag}, K., {Borgani}, S., {Murante}, G., \& {Springel}, V. 2009, \mnras, 399,
  497

\bibitem[{{Eisenstein} \& {Hut}(1998)}]{Eisenstein_1998}
{Eisenstein}, D.~J. \& {Hut}, P. 1998, \apj, 498, 137

\bibitem[{{Elahi} {et~al.}(2019){Elahi}, {Ca{\~n}as}, {Poulton}, {Tobar},
  {Willis}, {Lagos}, {Power}, \& {Robotham}}]{Elahi_2019}
{Elahi}, P.~J., {Ca{\~n}as}, R., {Poulton}, R. J.~J., {et~al.} 2019, \pasa, 36,
  e021

\bibitem[{{Elahi} {et~al.}(2011){Elahi}, {Thacker}, \& {Widrow}}]{Elahi_2011}
{Elahi}, P.~J., {Thacker}, R.~J., \& {Widrow}, L.~M. 2011, \mnras, 418, 320

\bibitem[{{Gill} {et~al.}(2004){Gill}, {Knebe}, \& {Gibson}}]{Gill_2004}
{Gill}, S. P.~D., {Knebe}, A., \& {Gibson}, B.~K. 2004, \mnras, 351, 399

\bibitem[{{Han} {et~al.}(2018){Han}, {Cole}, {Frenk}, {Benitez-Llambay}, \&
  {Helly}}]{Han_2018}
{Han}, J., {Cole}, S., {Frenk}, C.~S., {Benitez-Llambay}, A., \& {Helly}, J.
  2018, \mnras, 474, 604

\bibitem[{{Han} {et~al.}(2012){Han}, {Jing}, {Wang}, \& {Wang}}]{Han_2012}
{Han}, J., {Jing}, Y.~P., {Wang}, H., \& {Wang}, W. 2012, \mnras, 427, 2437

\bibitem[{{Hoffmann} {et~al.}(2014){Hoffmann}, {Planelles}, {Gazta{\~n}aga},
  {Knebe}, {Pearce}, {Lux}, {Onions}, {Muldrew}, {Elahi}, {Behroozi},
  {Ascasibar}, {Han}, {Maciejewski}, {Merchan}, {Neyrinck}, {Ruiz}, \&
  {Sgro}}]{Hoffmann_2014}
{Hoffmann}, K., {Planelles}, S., {Gazta{\~n}aga}, E., {et~al.} 2014, \mnras,
  442, 1197

\bibitem[{Hunter(2007)}]{Matplotlib}
Hunter, J.~D. 2007, Computing in Science \& Engineering, 9, 90

\bibitem[{{Iannuzzi} \& {Dolag}(2012)}]{Iannuzzi_2012}
{Iannuzzi}, F. \& {Dolag}, K. 2012, \mnras, 427, 1024

\bibitem[{{Ishiyama} {et~al.}(2021){Ishiyama}, {Prada}, {Klypin}, {Sinha},
  {Metcalf}, {Jullo}, {Altieri}, {Cora}, {Croton}, {de la Torre},
  {Mill{\'a}n-Calero}, {Oogi}, {Ruedas}, \&
  {Vega-Mart{\'\i}nez}}]{Ishiyama_2021}
{Ishiyama}, T., {Prada}, F., {Klypin}, A.~A., {et~al.} 2021, \mnras, 506, 4210

\bibitem[{{Kaviraj} {et~al.}(2017){Kaviraj}, {Laigle}, {Kimm}, {Devriendt},
  {Dubois}, {Pichon}, {Slyz}, {Chisari}, \& {Peirani}}]{Kaviraj_2017}
{Kaviraj}, S., {Laigle}, C., {Kimm}, T., {et~al.} 2017, \mnras, 467, 4739

\bibitem[{{Klypin} {et~al.}(1999){Klypin}, {Gottl{\"o}ber}, {Kravtsov}, \&
  {Khokhlov}}]{Klypin_1999}
{Klypin}, A., {Gottl{\"o}ber}, S., {Kravtsov}, A.~V., \& {Khokhlov}, A.~M.
  1999, \apj, 516, 530

\bibitem[{{Knebe} {et~al.}(2011){Knebe}, {Knollmann}, {Muldrew}, {Pearce},
  {Aragon-Calvo}, {Ascasibar}, {Behroozi}, {Ceverino}, {Colombi}, {Diemand},
  {Dolag}, {Falck}, {Fasel}, {Gardner}, {Gottl{\"o}ber}, {Hsu}, {Iannuzzi},
  {Klypin}, {Luki{\'c}}, {Maciejewski}, {McBride}, {Neyrinck}, {Planelles},
  {Potter}, {Quilis}, {Rasera}, {Read}, {Ricker}, {Roy}, {Springel}, {Stadel},
  {Stinson}, {Sutter}, {Turchaninov}, {Tweed}, {Yepes}, \& {Zemp}}]{Knebe_2011}
{Knebe}, A., {Knollmann}, S.~R., {Muldrew}, S.~I., {et~al.} 2011, \mnras, 415,
  2293

\bibitem[{{Knebe} {et~al.}(2013){Knebe}, {Pearce}, {Lux}, {Ascasibar},
  {Behroozi}, {Casado}, {Moran}, {Diemand}, {Dolag}, {Dominguez-Tenreiro},
  {Elahi}, {Falck}, {Gottl{\"o}ber}, {Han}, {Klypin}, {Luki{\'c}},
  {Maciejewski}, {McBride}, {Merch{\'a}n}, {Muldrew}, {Neyrinck}, {Onions},
  {Planelles}, {Potter}, {Quilis}, {Rasera}, {Ricker}, {Roy}, {Ruiz},
  {Sgr{\'o}}, {Springel}, {Stadel}, {Sutter}, {Tweed}, \& {Zemp}}]{Knebe_2013}
{Knebe}, A., {Pearce}, F.~R., {Lux}, H., {et~al.} 2013, \mnras, 435, 1618

\bibitem[{{Knollmann} \& {Knebe}(2009)}]{Knollmann_2009}
{Knollmann}, S.~R. \& {Knebe}, A. 2009, \apjs, 182, 608

\bibitem[{{Martin-Alvarez} {et~al.}(2017){Martin-Alvarez}, {Planelles}, \&
  {Quilis}}]{Martin-Alvarez_2017}
{Martin-Alvarez}, S., {Planelles}, S., \& {Quilis}, V. 2017, \apss, 362, 91

\bibitem[{{Murante} {et~al.}(2010){Murante}, {Monaco}, {Giovalli}, {Borgani},
  \& {Diaferio}}]{Murante_2010}
{Murante}, G., {Monaco}, P., {Giovalli}, M., {Borgani}, S., \& {Diaferio}, A.
  2010, \mnras, 405, 1491

\bibitem[{{Navarro} {et~al.}(1997){Navarro}, {Frenk}, \& {White}}]{NFW}
{Navarro}, J.~F., {Frenk}, C.~S., \& {White}, S. D.~M. 1997, \apj, 490, 493

\bibitem[{{Navarro-Gonz{\'a}lez} {et~al.}(2013){Navarro-Gonz{\'a}lez},
  {Ricciardelli}, {Quilis}, \& {Vazdekis}}]{Navarro_2013}
{Navarro-Gonz{\'a}lez}, J., {Ricciardelli}, E., {Quilis}, V., \& {Vazdekis}, A.
  2013, \mnras, 436, 3507

\bibitem[{Oliphant(2006)}]{Numpy}
Oliphant, T.~E. 2006, {A guide to NumPy}, Vol.~1 (Trelgol Publishing USA)

\bibitem[{{Onions} {et~al.}(2013){Onions}, {Ascasibar}, {Behroozi}, {Casado},
  {Elahi}, {Han}, {Knebe}, {Lux}, {Merch{\'a}n}, {Muldrew}, {Neyrinck}, {Old},
  {Pearce}, {Potter}, {Ruiz}, {Sgr{\'o}}, {Tweed}, \& {Yue}}]{Onions_2013}
{Onions}, J., {Ascasibar}, Y., {Behroozi}, P., {et~al.} 2013, \mnras, 429, 2739

\bibitem[{{Onions} {et~al.}(2012){Onions}, {Knebe}, {Pearce}, {Muldrew}, {Lux},
  {Knollmann}, {Ascasibar}, {Behroozi}, {Elahi}, {Han}, {Maciejewski},
  {Merch{\'a}n}, {Neyrinck}, {Ruiz}, {Sgr{\'o}}, {Springel}, \&
  {Tweed}}]{Onions_2012}
{Onions}, J., {Knebe}, A., {Pearce}, F.~R., {et~al.} 2012, \mnras, 423, 1200

\bibitem[{{Pillepich} {et~al.}(2018){Pillepich}, {Springel}, {Nelson}, {Genel},
  {Naiman}, {Pakmor}, {Hernquist}, {Torrey}, {Vogelsberger}, {Weinberger}, \&
  {Marinacci}}]{Pillepich_2018}
{Pillepich}, A., {Springel}, V., {Nelson}, D., {et~al.} 2018, \mnras, 473, 4077

\bibitem[{{Pillepich} {et~al.}(2014){Pillepich}, {Vogelsberger}, {Deason},
  {Rodriguez-Gomez}, {Genel}, {Nelson}, {Torrey}, {Sales}, {Marinacci},
  {Springel}, {Sijacki}, \& {Hernquist}}]{Pillepich_2014}
{Pillepich}, A., {Vogelsberger}, M., {Deason}, A., {et~al.} 2014, \mnras, 444,
  237

\bibitem[{{Planelles} {et~al.}(2018){Planelles}, {Mimica}, {Quilis}, \&
  {Cuesta-Mart{\'\i}nez}}]{Planelles_2018}
{Planelles}, S., {Mimica}, P., {Quilis}, V., \& {Cuesta-Mart{\'\i}nez}, C.
  2018, \mnras, 476, 4629

\bibitem[{{Planelles} \& {Quilis}(2010)}]{Planelles_2010}
{Planelles}, S. \& {Quilis}, V. 2010, \aap, 519, A94

\bibitem[{{Planelles} \& {Quilis}(2013)}]{Planelles_2013}
{Planelles}, S. \& {Quilis}, V. 2013, \mnras, 428, 1643

\bibitem[{{Planelles} {et~al.}(2015){Planelles}, {Schleicher}, \&
  {Bykov}}]{Planelles_2015}
{Planelles}, S., {Schleicher}, D.~R.~G., \& {Bykov}, A.~M. 2015, \ssr, 188, 93

\bibitem[{{Press} \& {Schechter}(1974)}]{Press_1974}
{Press}, W.~H. \& {Schechter}, P. 1974, \apj, 187, 425

\bibitem[{{Quilis}(2004)}]{Quilis_2004}
{Quilis}, V. 2004, \mnras, 352, 1426

\bibitem[{{Quilis} {et~al.}(2020){Quilis}, {Mart{\'\i}}, \&
  {Planelles}}]{Quilis_2020}
{Quilis}, V., {Mart{\'\i}}, J.-M., \& {Planelles}, S. 2020, \mnras, 494, 2706

\bibitem[{{Quilis} {et~al.}(2017){Quilis}, {Planelles}, \&
  {Ricciardelli}}]{Quilis_2017}
{Quilis}, V., {Planelles}, S., \& {Ricciardelli}, E. 2017, \mnras, 469, 80

\bibitem[{{Schaye} {et~al.}(2015){Schaye}, {Crain}, {Bower}, {Furlong},
  {Schaller}, {Theuns}, {Dalla Vecchia}, {Frenk}, {McCarthy}, {Helly},
  {Jenkins}, {Rosas-Guevara}, {White}, {Baes}, {Booth}, {Camps}, {Navarro},
  {Qu}, {Rahmati}, {Sawala}, {Thomas}, \& {Trayford}}]{Schaye_2015}
{Schaye}, J., {Crain}, R.~A., {Bower}, R.~G., {et~al.} 2015, \mnras, 446, 521

\bibitem[{{Schechter}(1976)}]{Schechter_1976}
{Schechter}, P. 1976, \apj, 203, 297

\bibitem[{{Springel}(2010)}]{Springel_2010}
{Springel}, V. 2010, \mnras, 401, 791

\bibitem[{{Springel} {et~al.}(2001){Springel}, {White}, {Tormen}, \&
  {Kauffmann}}]{Springel_2001}
{Springel}, V., {White}, S. D.~M., {Tormen}, G., \& {Kauffmann}, G. 2001,
  \mnras, 328, 726

\bibitem[{{Tinker} {et~al.}(2008){Tinker}, {Kravtsov}, {Klypin}, {Abazajian},
  {Warren}, {Yepes}, {Gottl{\"o}ber}, \& {Holz}}]{Tinker_2008}
{Tinker}, J., {Kravtsov}, A.~V., {Klypin}, A., {et~al.} 2008, \apj, 688, 709

\bibitem[{{Tormen} {et~al.}(2004){Tormen}, {Moscardini}, \&
  {Yoshida}}]{Tormen_2004}
{Tormen}, G., {Moscardini}, L., \& {Yoshida}, N. 2004, \mnras, 350, 1397

\bibitem[{{Vall{\'e}s-P{\'e}rez} {et~al.}(2020){Vall{\'e}s-P{\'e}rez},
  {Planelles}, \& {Quilis}}]{Valles_2020}
{Vall{\'e}s-P{\'e}rez}, D., {Planelles}, S., \& {Quilis}, V. 2020, \mnras, 499,
  2303

\bibitem[{{Vall{\'e}s-P{\'e}rez} {et~al.}(2021){Vall{\'e}s-P{\'e}rez},
  {Planelles}, \& {Quilis}}]{Valles_2021}
{Vall{\'e}s-P{\'e}rez}, D., {Planelles}, S., \& {Quilis}, V. 2021, \mnras, 504,
  510

\bibitem[{{Villaescusa-Navarro} {et~al.}(2021){Villaescusa-Navarro},
  {Angl{\'e}s-Alc{\'a}zar}, {Genel}, {Spergel}, {Somerville}, {Dave},
  {Pillepich}, {Hernquist}, {Nelson}, {Torrey}, {Narayanan}, {Li}, {Philcox},
  {La Torre}, {Maria Delgado}, {Ho}, {Hassan}, {Burkhart}, {Wadekar},
  {Battaglia}, {Contardo}, \& {Bryan}}]{Villaescusa-Navarro_2021}
{Villaescusa-Navarro}, F., {Angl{\'e}s-Alc{\'a}zar}, D., {Genel}, S., {et~al.}
  2021, \apj, 915, 71

\bibitem[{{Villaescusa-Navarro} {et~al.}(2022){Villaescusa-Navarro}, {Genel},
  {Angl{\'e}s-Alc{\'a}zar}, {Perez}, {Villanueva-Domingo}, {Wadekar}, {Shao},
  {Mohammad}, {Hassan}, {Moser}, {Lau}, {Machado Poletti Valle}, {Nicola},
  {Thiele}, {Jo}, {Philcox}, {Oppenheimer}, {Tillman}, {Hahn}, {Kaushal},
  {Pisani}, {Gebhardt}, {Delgado}, {Caliendo}, {Kreisch}, {Wong}, {Coulton},
  {Eickenberg}, {Parimbelli}, {Ni}, {Steinwandel}, {La Torre}, {Dave},
  {Battaglia}, {Nagai}, {Spergel}, {Hernquist}, {Burkhart}, {Narayanan},
  {Wandelt}, {Somerville}, {Bryan}, {Viel}, {Li}, {Irsic}, {Kraljic}, \&
  {Vogelsberger}}]{Villaescusa-Navarro_2022}
{Villaescusa-Navarro}, F., {Genel}, S., {Angl{\'e}s-Alc{\'a}zar}, D., {et~al.}
  2022, arXiv e-prints, arXiv:2201.01300

\bibitem[{{Virtanen} {et~al.}(2020){Virtanen}, {Gommers}, {Oliphant},
  {Haberland}, {Reddy}, {Cournapeau}, {Burovski}, {Peterson}, {Weckesser},
  {Bright}, {van der Walt}, {Brett}, {Wilson}, {Jarrod Millman}, {Mayorov},
  {Nelson}, {Jones}, {Kern}, {Larson}, {Carey}, {Polat}, {Feng}, {Moore},
  {VanderPlas}, {Laxalde}, {Perktold}, {Cimrman}, {Henriksen}, {Quintero},
  {Harris}, {Archibald}, {Ribeiro}, {Pedregosa}, {van Mulbregt}, \&
  {Contributors}}]{Scipy}
{Virtanen}, P., {Gommers}, R., {Oliphant}, T.~E., {et~al.} 2020, Nature
  Methods, 17, 261

\bibitem[{{Vogelsberger} {et~al.}(2014){Vogelsberger}, {Genel}, {Springel},
  {Torrey}, {Sijacki}, {Xu}, {Snyder}, {Bird}, {Nelson}, \&
  {Hernquist}}]{Vogelsberger_2014}
{Vogelsberger}, M., {Genel}, S., {Springel}, V., {et~al.} 2014, \nat, 509, 177

\bibitem[{{Vogelsberger} {et~al.}(2020){Vogelsberger}, {Marinacci}, {Torrey},
  \& {Puchwein}}]{Vogelsberger_2020}
{Vogelsberger}, M., {Marinacci}, F., {Torrey}, P., \& {Puchwein}, E. 2020,
  Nature Reviews Physics, 2, 42

\bibitem[{{Weinberger} {et~al.}(2020){Weinberger}, {Springel}, \&
  {Pakmor}}]{Weinberger_2020}
{Weinberger}, R., {Springel}, V., \& {Pakmor}, R. 2020, \apjs, 248, 32

\end{thebibliography}

\begin{appendix}

\section{Solution of Poisson's equation in spherical symmetry}
\label{s.appendix.unbinding}
The local unbinding procedure, described in Sect. \ref{s:algorithm.particles}, relies on the computation of the gravitational potential in spherical symmetry. Here we describe our implementation of the procedure.

In general, the gravitational potential, $\Phi$, of any -continuous or discrete- density distribution $\rho$ is obtained by solving Poisson's equation,

\begin{equation}
\nabla_{\vec{r}}^2 \Phi = \frac{1}{a^2} \nabla_{\vec{x}}^2 \Phi = 4 \pi G \rho,
\end{equation}

\noindent which is an elliptic partial derivative equation, where $\vec{r}$ and $\vec{x} \equiv \vec{r}/a$ are, respectively, the physical and the comoving position vectors, related by the expansion factor $a(t)$ which solves Friedmann equations for a given cosmology. Under the assumption of spherical symmetry in the density distribution (and thus in the potential), the Poisson equation reduces to a non-linear second-order ordinary differential equation,

\begin{equation}
 \dv{}{r}\left[ r^2 \dv{\Phi}{r} \right] = 4 \pi G \rho(r) r^2.  
\end{equation}

If the mass distribution is bound, it is always possible to take $\lim_{r\to \infty} \Phi(r) = 0$ and show that

\begin{equation}
\Phi(r) = \int_{\infty}^{r} \frac{G M(<r')}{r'^2} \dd r',
\end{equation}

\noindent being $M(<r) \equiv \int_{0}^{r} 4\pi r'^2 \rho(r') \dd r'$ the mass enclosed in a sphere of radius $r$. Assuming the mass distribution to be bound within a maximum radius $r_\mathrm{max}$, is then straightforward to rewrite this expression to a more convenient form for its numerical integration:

\begin{equation}
    \Phi(r) = \int_0^r \frac{G M(<r')}{r'^2} \dd r' - \frac{G M(<r_\mathrm{max})}{r_\mathrm{max}} - \int_0^{r_\mathrm{max}} \frac{G M(<r')}{r'^2} \dd r'
    \label{eq:poisson_integral}
\end{equation}

Note that only the first term in Eq. \ref{eq:poisson_integral} is a function of $r$. Operationally, we use the following recipe for efficiently performing the integration.

\begin{enumerate}
    \item Start from a list of $n_\mathrm{part}$ particles ($i=1,\, \dots, \,n_\mathrm{part}$), with masses $\{m_i\}_{i=1}^{n_\mathrm{part}}$, sorted in increasing comoving distance to the halo centre (density peak),  $\{r_i \; / \; r_j \leq r_k \forall j \leq k\}_{i=1}^{n_\mathrm{part}}$. Denote $r_\mathrm{max} = r_\mathrm{n_\mathrm{part}}$.
    \item Compute the cummulative mass profile, which we shall denote $M_i \equiv \sum_{j=1}^i m_j$.
    \item Identify $\mathcal{J}$ as the smallest integer such that $r_\mathcal{J} > 0.01 r_\mathrm{max}$, and define $\tilde \Phi_{j \leq \mathcal{J}} = \frac{M_\mathcal{J}}{r_\mathcal{J}}$.
    \item Compute iteratively, for $j=\mathcal{J}+1,\, \dots, \,n_\mathrm{part}$, the value of $\tilde \Phi$ at the position of the next particle as $\tilde \Phi_j = \tilde \Phi_{j-1} + \frac{M_j (r_j - r_{j-1})}{r_j^2}$.
    \item Define the constant $C = \tilde \Phi_{n_\mathrm{part}} + \frac{M_{n_\mathrm{part}}}{r_{n_\mathrm{part}}}$.
    \item The spherically-symmetric gravitational potential at comoving radius $r_i$ is $\Phi_i = \frac{G}{a}(C - \tilde \Phi_i) \leq 0 \;\forall \; i$.
\end{enumerate}

Note that we perform step 3, i.e. assuming a constant gravitational potential in the inner $1\%$ of the radial space, to mitigate numerical errors since $r_i$ can be arbitrarily small. This has a negligible impact on the unbinding procedure, since it is rare to find unbound particles near halo centres.

\section{Computation of the gravitational binding energy by sampling}
\label{s.appendix.gravitational_energy}

\begin{figure}
    \centering
    \includegraphics[width=\linewidth]{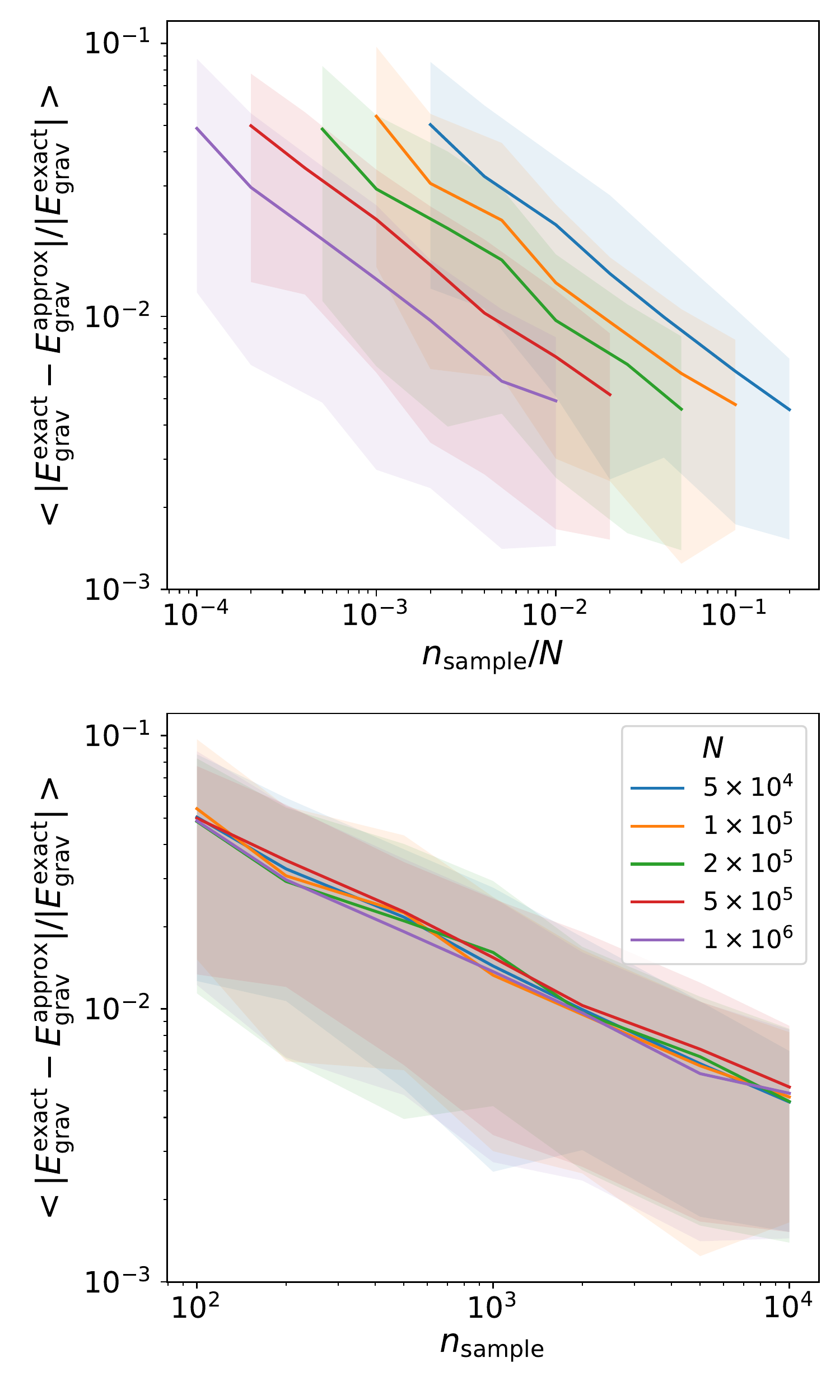}
    \caption{Dependence of the error in obtaining the gravitational binding energy by sampling, rather than by direct summation over all the pairs of particles, with the fraction $n_\mathrm{sample}/N$ (\textit{upper panel}) and with just the number of sample particles $n_\mathrm{sample}$ (\textit{lower panel}). Note that the number of pairs of particles evaluated is $n_\mathrm{sample}^2$. Different colours, according to the legend in the lower panel, represent different number of particles within the halo. The shadowed regions correspond to $(16-84)$\% confidence regions.}
    \label{figB1}
\end{figure}

The gravitational binding (or potential) energy of a system of $N$ particles is explicitly given by the sum over all particle pairs:

\begin{equation}
    E_\mathrm{grav} = - G \sum_{i=1}^N \sum_{j=i+1}^N \frac{m_i m_j}{|\vec{r_i} - \vec{r_j}|},
    \label{eq:grav_energy_direct_sum}
\end{equation}

\noindent which therefore implies and $\mathcal{O}(N^2)$ calculation, getting prohibitively expensive for large enough $N$. In \ASOHFns, we perform a direct summation over the $N(N-1)/2$ pairs for haloes with $N\leq N_\mathrm{max}^\mathrm{energy}$ particles (see the discussion of Fig. \ref{figB1} for the dependence of the accuracy of the results on this parameter), while for larger haloes we use a sampling estimate of this quantity, which we describe and test below.

In order to estimate the gravitational binding energy of large haloes by sampling, we consider $n_\mathrm{sample}^2 = \max(\lfloor N_\mathrm{max}^\mathrm{energy}/\sqrt{2} \rfloor, \, 0.01 N)^2$ pairs of particles and compute the contribution of each to the potential energy, i.e. $E_\mathrm{grav,ij} = -G \frac{m_i m_j}{|\vec{r_i} - \vec{r_j}|}$ for the pair of particles $(ij)$. The procedure is as follows:

\begin{itemize}
    \item We randomly pick $n_\mathrm{sample}$ particles (with replacement), which we shall denote $i=1,\, \dots , \, n_\mathrm{sample}$.
    \item For each of these particles, we pick a new set of $n_\mathrm{sample}$ particles (with replacement), which we shall denote $j=1,\, \dots , \, n_\mathrm{sample}$, and compute, for particle $i$, its contribution to the gravitational energy, $E_\mathrm{grav,i}^\mathrm{sample} = \sum_{j=1}^{n_\mathrm{sample}} E_\mathrm{grav,ij}$.
    \item The gravitational energy per pair of particles will therefore be
    \begin{equation}
    \begin{aligned}
        \langle E_\mathrm{grav,pairs} \rangle &= \frac{1}{n_\mathrm{sample}^2} \sum_{i=1}^{n_\mathrm{sample}} E_\mathrm{grav,i}^\mathrm{sample} = \frac{1}{n_\mathrm{sample}^2} \sum_{i=1}^{n_\mathrm{sample}} \sum_{j=1}^{n_\mathrm{sample}} E_\mathrm{grav,ij}.
    \end{aligned}
    \end{equation}
    \item Since there are $N(N-1)/2$ pairs of particles in the halo, we estimate its gravitational energy as
    \begin{equation}
        E_\mathrm{grav} \approxeq \frac{N(N-1)}{2} \langle E_\mathrm{grav,pairs} \rangle.
    \end{equation}
    Note that $E_\mathrm{grav} \propto \langle E_\mathrm{grav,pairs} \rangle$ and thus the relative error in $E_\mathrm{grav}$ is equal to the relative error in $\langle E_\mathrm{grav,pairs} \rangle$, which is in turn proportional to $1/\sqrt{n_\mathrm{sample}}$ and independent of $N$. This is graphically exemplified in Fig. \ref{figB1}.
    \item The gravitational binding energy per unit mass of particle $i$ is proportional to $E_\mathrm{grav,i}/m_i$. Thus, we report as the most bound particle the particle with the most negative value of $E_\mathrm{grav,i}/m_i$, which incidentally can be used as an estimation of the location of the minimum of gravitational potential. The uncertainty in determining this position can be estimated as the sphere, with centre in the most-bound particle, which encloses $\lceil N/n_\mathrm{sample} \rceil$ particles.
\end{itemize}

To assess the convergence of this method, we have generated NFW haloes with concentration arbitrarily fixed to $c=10$, and radius $R=1.5$ (in arbitrary units), realised using different numbers of particles, $N$, from $5\times 10^4$ to $10^6$. We have then computed the gravitational energy by direct, $\mathcal{O} (N^2)$ summation as in Eq. \ref{eq:grav_energy_direct_sum}; and by the sampling methods varying $n_\mathrm{sample}$ between $10^2$ and $10^4$. For each pair of values of ($N$, $n_\mathrm{sample}$), we have performed $n_\mathrm{boots}=100$ bootstrap iterations of the calculation by sampling, so that we can estimate the mean relative error in the gravitational potential energy comparing with the result by direct summation, and its $(16-84)\%$ confidence intervals.

This is shown in Fig. \ref{figB1}, whose upper panel shows the mean of the relative error, computed over the 100 bootstrap iterations, in computing the gravitational energy by sampling. Each colour represents a different number of particles in the halo according to the legend in the lower panel, while in the horizontal axis we show the fraction of $n_\mathrm{sample}$ to $N$. It can be easily seen that, for each $N$, the error decreases as $1/\sqrt{n_\mathrm{sample}}$. These curves can be brought to match each other if we represent the relative error as a function of $n_\mathrm{sample}$, as shown in the lower panel of Fig. \ref{figB1}. In this case, we see that, regardless the number of particles of the halo, mean errors fall below $1\%$ when using more than $n_\mathrm{sample} \approx 2000$, while $n_\mathrm{sample}$ must be increased to $\sim 5000-6000$ to achieve an error below $1\%$ at the 84-percentile. 

It would be reasonable to ask why the error scales as $1/\sqrt{n_\mathrm{sample}}$, instead of $1/\sqrt{n_\mathrm{pairs}}=1/n_\mathrm{sample}$, which would be expected from averaging $n_\mathrm{pairs}$ values of the energy per pair of particles. This is due to the fact that we are considering $n_\mathrm{sample}$ particles, and estimating the gravitational binding energy of each of these with a new set of $n_\mathrm{sample}$ particles. This is fundamentally different from choosing $n_\mathrm{pairs}$ completely independent pairs, in which case it is easy to check that the error indeed scales as $\propto 1/n_\mathrm{sample}$. However, this latter method, although much faster in terms of convergence (lower values of $n_\mathrm{sample}$ are required and, therefore, the computational cost is greatly reduced), does not allow us to estimate the most-bound particles of haloes (and, therefore, the position of the gravitational potential minima). This is so because, for this aim, we require a good estimate of the energies of a subsample of individual particles. This cannot be achieved by randomly picking pairs of particles, since each particle will be considered a few times, as much. This is the reason why, in \ASOHFns, we opt for this slower-convering method.

\end{appendix}

\end{document}